\documentclass{cernyrep} 
\usepackage{wrapfig}
\usepackage[bookmarks, colorlinks=true, linktoc=page, linkcolor=black, citecolor=black, urlcolor=blue]{hyperref}
\usepackage{fancyhdr}
\fancyhfoffset{4 mm}
\fancypagestyle{ARTTITLE}{%
\fancyhf{} 
\lhead{\small{Proceedings of the 2018 CERN--Accelerator--School course on \it{Beam Instrumentation}, Tuusula, (Finland)}} 
\lfoot{Available online at \url{https://cas.web.cern.ch/previous-schools}}
\rfoot{\thepage\hspace*{3mm}}
 
}
\begin{document}
\title{Analog electronics for beam instrumentation}
\author{J. Belleman}
\institute{CERN, Geneva, Switzerland}

\begin{abstract}
The task of analog front-end electronics in beam instrumentation is
to optimize the useful information content of the signal delivered
by an instrument. It must suppress signal components that do
not contribute to the measured quantity. It must filter to put bounds
on bandwidth and possibly dynamic range, to relax the demands made
of subsequent processing stages.  It must minimize noise, reject
interference and match the signal to transmission media and digital
acquisition equipment.  Since the circuitry must often operate in
radio-active areas, the accent is on passive electronics.

\end{abstract}

\maketitle 
\thispagestyle{ARTTITLE}
\section{Introduction}

It is tempting to just connect
a measuring device to an ADC and to rely entirely on digital signal processing
to extract useful information. Most often, this would be far
from optimal.  The signals produced by most transducers are a poor
match to any ADC.  To make the most of the available signal energy,
analog signal conditioning is crucial. No amount of
digital processing can recover information that isn't there anymore.
Analog electronics provide rejection of interference, filtering and
amplification.

I can't hope to give an anywhere near complete view of the vast
subject of analog electronics. My purpose is rather to give you a
taste of some useful subjects, with some hints, references and relevant
keywords, so that you can look up more rigorous treatment elsewhere,
should you need to do so.  For a more comprehensive treatment of
many subjects, the primary reference would be the excellent 'Art of
Electronics'~\cite{bib:AoE}. Other useful references on electronics and
instrument design are \cite{bib:Williams} and \cite{bib:Hobbs}. In this
text, I chose to treat some subjects that I felt are somewhat neglected
in most general treatises on electronics, but which are nevertheless
significant for beam instrumentation.

Because beam instrumentation must often work in radio-active environments,
where active semiconductor electronics may not survive for very long, 
much of what will follow concerns passive electronics.

\section{Transmission lines}
Transmission lines serve to carry signals from one place to another,
of course, but are also used in transformers, splitters and
combiners, hybrids and directional couplers, pulse forming networks, 
filters and resonators, bias and matching networks and more.
A few of those applications, relevant to beam instrumentation, 
will be shown in what follows.

\begin{wrapfigure}{R}{0.3\textwidth}
 \center\includegraphics[width=0.75\linewidth]{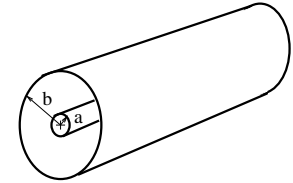}
 \caption{Coaxial conductors}
 \label{fig:coax}
\end{wrapfigure}
The most common transmission line is the coaxial cable
(Fig.~\ref{fig:coax}), a central wire completely enclosed in a conductive
shield over its full length.  The shield prevents the signal from
radiating away and protects it from external fields.  Such a cable is
characterized by a distributed parallel capacitance and series inductance,
determined by the geometry and by the materials used to make the cable.
The capacitance, in farad/meter, can be found by applying Gauss' law in the space between
the conductors:
\begin{equation}
C_0 = \frac{1}{\int_{a}^{b}\frac{1}{2\pi\varepsilon r}dr} = \frac{2\pi\varepsilon}{\ln{\frac{b}{a}}},
\end{equation}
with $\varepsilon = \varepsilon_0 \varepsilon_r$ the electric permittivity
of the insulator in the space between the conductors.
Similarly we can get the inductance in henry/meter using Amp\`ere's law:
\begin{equation}
L_0 = \int_{a}^{b}\frac{\mu}{2\pi r}dr = \frac{\mu}{2\pi}\ln{\frac{b}{a}},
\end{equation}
with $\mu = \mu_0 \mu_r$ the magnetic permeability of the material inside.
The resultant impedance seen by a signal source connected to an endless cable
is then
\begin{equation}
Z_0 = \sqrt{\frac{L_0}{C_0}}.
\label{eq:Z0}
\end{equation}
$Z_0$ is the \emph{characteristic impedance} of the cable. Note that
it's a real value, a pure resistance. 
Commonly available coax can be had with impedances ranging from
about 10 to 190~$\Omega$, but the most common values are 50~$\Omega$
or 75~$\Omega$.
For $Z_0$~=~50~$\Omega$, typically
$C_0$~=~100~pF/m and $L_0$~=~250~nH/m.

Practical cables also have a distributed series resistance
$R_0$ and a parallel (leakage) conductance $G_0$.  Incorporating those into
the expression for $Z_0$ results in
\begin{equation}
Z_0 = \sqrt{\frac{j \omega L_0 + R_0}{j \omega C_0 + G_0}},
\label{eq:tline}
\end{equation}
which, unless $L_0/C_0 = R_0/G_0$, is now a complex value dependent on
frequency. Often the leakage conductance $G_0$ is negligible, but the
series resistance $R_0$ is not.  These additional elements are responsible
for transmission loss.

Although Eq.~(\ref{eq:tline}) doesn't say so, the values of $R_0$ and $G_0$
are not constant. Both depend on frequency. $R_0$ is modified by the
\emph{skin effect}, the tendency of high-frequency AC currents to flow
in a thin surface layer. $G_0$ is affected by dielectric losses.
In a certain frequency range, generally from about 100~kHz to several
gigahertz, the wire
resistance and leakage conductance hardly affect the cable impedance
which then converges on the value of Eq.~(\ref{eq:Z0}).
Cable loss expressed in decibels/m increases with the square root of
frequency, which is pretty damning if you need to transport
high-frequency signals over long distances.


\subsection{Signal propagation}
A signal in a coaxial transmission line can be thought of as
an electromagnetic wave propagating through the space between the
conductors. The wave is accompanied by co-moving charges in the conductors
that compose the cable. The charge on the central conductor is mirrored
by an opposite charge flowing along the inner surface of the screen.
As a result, there are no magnetic or electric fields outside of
the screen. This is the \emph{normal} or \emph{odd} mode of signal
propagation.  
The signal propagation velocity in the cable is
\begin{equation}
v_0 = \sqrt{\frac{1}{L_0 C_0 }}.
\end{equation}
In the absence of any dielectric material in the cable, this is equal
to $c$.
Practical cables of course need some kind of insulating support for
the central conductor,  so $\varepsilon_r > 1$ and the signal propagation
velocity is then reduced to $c / \sqrt{\varepsilon_r}$, usually somewhere
in the range of $0.6~c$ to $0.9~c$, depending on the chosen dielectric and
its consistency: solid, foam or discrete spacers.
Usually, cables do not contain any magnetic materials,
so $\mu_r = 1$.

It's also possible to have a current flowing over the outside of the
screen. This is the \emph{common} or \emph{even} mode.  For that mode,
the fields are in the space around the cable, with nothing inside.
Whereas the odd mode has a well defined characteristic impedance $Z_0$,
the even mode impedance is affected by the surroundings of the cable
and is hard to predict.  Even mode current is undesired and one seeks to
reduce it as much as possible. This can be done, for example, by inserting
transformers or baluns, sometimes simply by slipping a ferrite ring over
the cable. More on this later.

\subsection{Other forms of transmission lines}
A transmission line doesn't have to be coaxial. Whenever there is a
convenient nearby conductor to support an image current, we have
a transmission line. Commonly seen forms are twisted pairs, PCB tracks
over full copper planes (stripline or microstrip), or parallel wires
or PCB tracks (e.g. twin-lead, coplanar waveguide). For some geometries,
a closed formula can give the characteristic impedance $Z_0$ and for
others, empirical approximated expressions exist. Many software tools
exist to help, from very simple and free (e.g. atlc), to sophisticated
EM field simulators (CST Microwave Studio, HFSS ,...).

\begin{center}
 \begin{figure}[h]
  \begin{minipage}[c]{0.49\linewidth} 
   \center\includegraphics[width=0.8\linewidth]{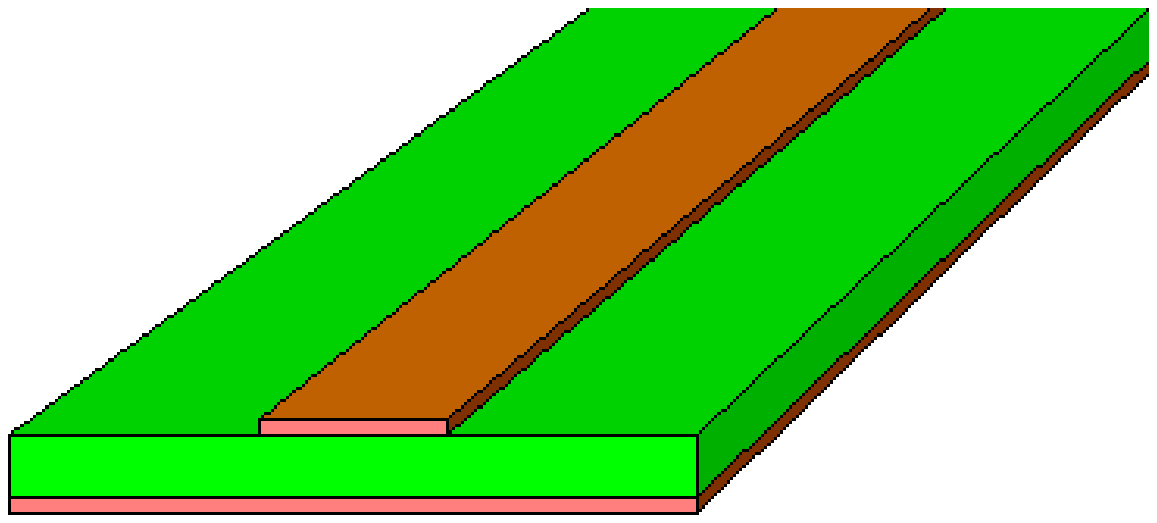}
   \caption{Microstrip}
   \label{fig:microstrip}
  \end{minipage}
  \begin{minipage}[c]{0.49\linewidth} 
   \center\includegraphics[width=0.8\linewidth]{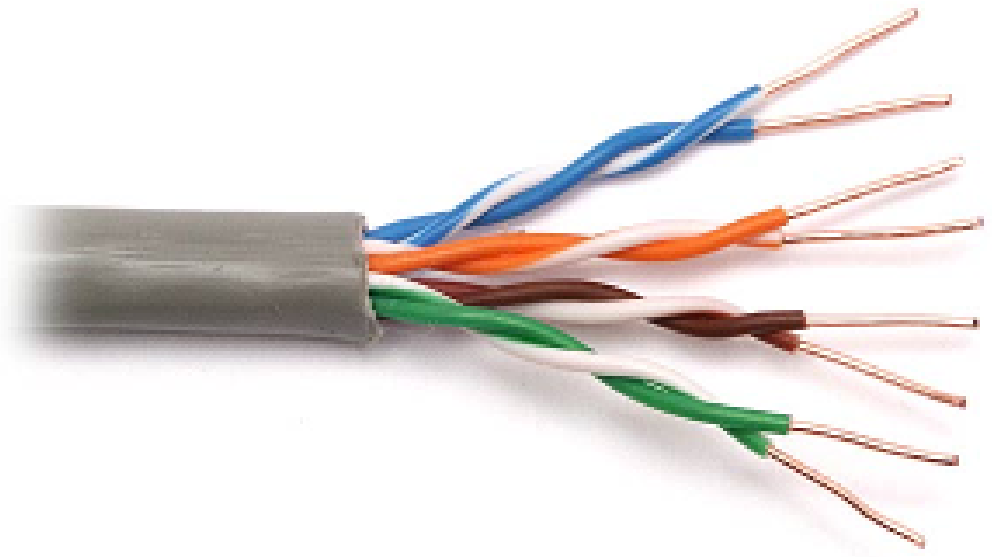}
   \caption{Twisted pair wire}
   \label{fig:tp}
  \end{minipage}
 \end{figure}
\end{center}

\subsection{Reflection and termination}
Any discontinuity encountered by the propagating wave in a
transmission line structure will
cause a backward travelling reflection.
The superposition of forward and
backward travelling waves gives rise to standing waves,
like vibrations in a taut string.
However, if a finite length of
transmission line is terminated with a resistance of value $Z_0$, 
no reflections will occur, since to the incident wave this
just looks like the continuation of the transmission line.

\begin{wrapfigure}{R}{0.5\textwidth}
\center\includegraphics[width=0.8\linewidth]{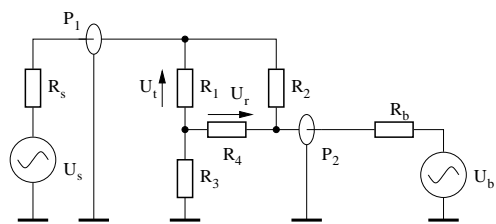}
\caption{Simple Wheatstone bridge directional coupler}
\label{fig:NA3}                
\end{wrapfigure}               
It is possible to distinguish forward and backward travelling
waves using directional couplers. A simple example of such
a coupler is the Wheatstone bridge (Fig.~\ref{fig:NA3}).
Assume all resistors are the same value. Consider a signal
injected into port $P_1$ by source $U_s$, with source $U_b$ zero.
Since $R_1 / R_3 = R_2 / R_b$, no signal will appear across $R_4$.
Likewise, if we now inject a signal from source $U_b$ into
port $P_2$, with source $U_s$ zero, since $R_2 / R_s = R_4 / R_3$,
no signal will appear across $R_1$. So across $R_1$, we see
only signal travelling right due to $U_s$, and across $R_4$, we
see only signal travelling left due to $U_b$. 
A circuit like this is at the heart of radio-frequency network
analyzers. 

\begin{center}
 \begin{figure}[h]
  \begin{minipage}[c]{0.49\linewidth} 
   \center\includegraphics[width=0.9\linewidth]{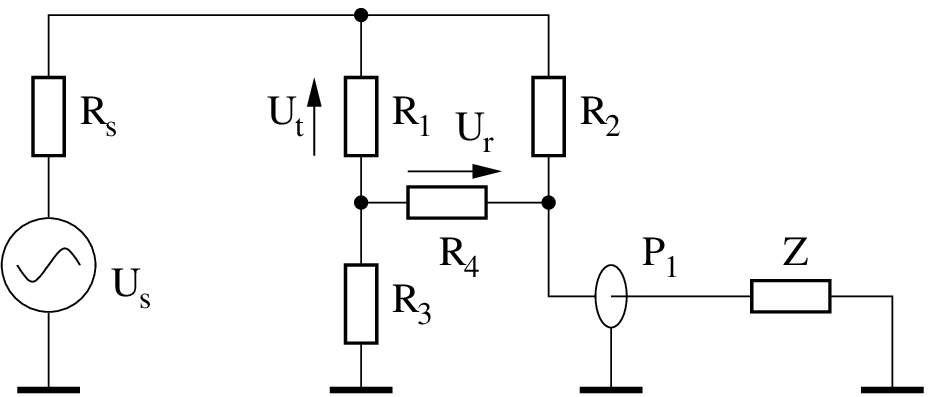}
   \caption{Network analyzer impedance measurement}
   \label{fig:NA}
  \end{minipage}
  \begin{minipage}[c]{0.49\linewidth} 
   \center\includegraphics[width=0.9\linewidth]{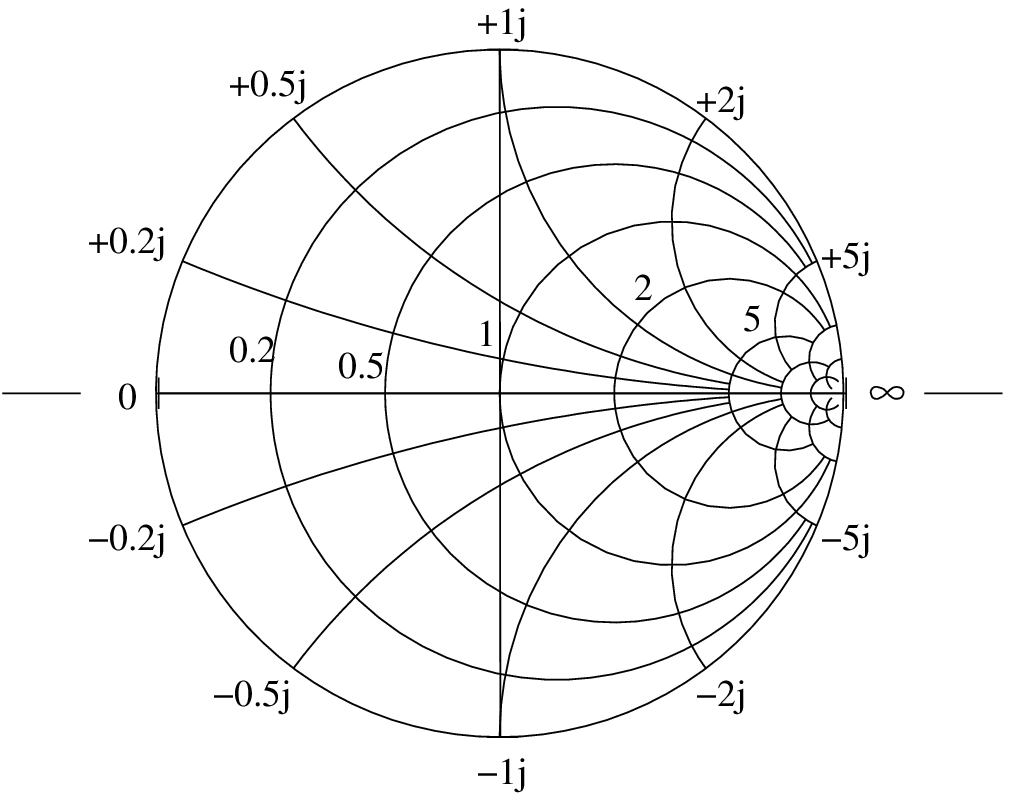}
   \caption{Smith Chart}
   \label{fig:smithchart}
  \end{minipage}
 \end{figure}
\end{center}

A slightly different view, but which some reflection will show to be
equivalent if $U_s$ and $U_b$ were coherent sinusoids, is the situation
depicted in figure~\ref{fig:NA}. Instead of an independent source $U_b$ with
fixed source impedance $R_b$, we substitute a variable impedance $Z$.

Any value of $Z$ different from the other resistors will 
cause a reflection $U_r$ appearing across $R_4$, in the same way as
the backward travelling signal of figure~\ref{fig:NA3}. 
\begin{equation}
 \frac{U_r}{U_s} = \frac{Z-R}{8(Z+R)}
 \label{eq:bilinear}
\end{equation}
The bilinear transformation Eq.~(\ref{eq:bilinear}), without the factor
$\frac{1}{8}$, is known as the \emph{reflection coefficient} and it maps any
passive complex impedance Z into a circle, with the perfect termination
value in the centre.  As is done in figure~\ref{fig:smithchart}, we
can draw new grid lines that allow us to read off the (normalized)
value of Z directly from the value of $U_r/U_s$.  The grid lines are
loci of constant real or imaginary parts of the impedance $Z$ and are
all circles.  This is known as a Smith chart \cite{bib:smith1939}. 
Before the generalized availability of computers, this chart
was so useful as a graphical calculation aid that stylized pictures
of the chart's grid became trademarks for many companies involved in
radio-frequency signal processing.

\subsection{Time domain reflectometry}
A very useful technique to sound the properties of transmission line
structures, cables, connectors, vacuum feedthroughs and more is Time
Domain Reflectometry (TDR). A fast step is launched into a transmission
line structure, and the reflections, the echos, return information about
what's inside. The technique is eminently useful to locate faults in
cables and connectors, yielding information about the location and the
nature of the fault.

\begin{figure}[h]
 \begin{center}
  \begin{minipage}[c]{0.49\linewidth}
   \center\includegraphics[width=0.8\linewidth]{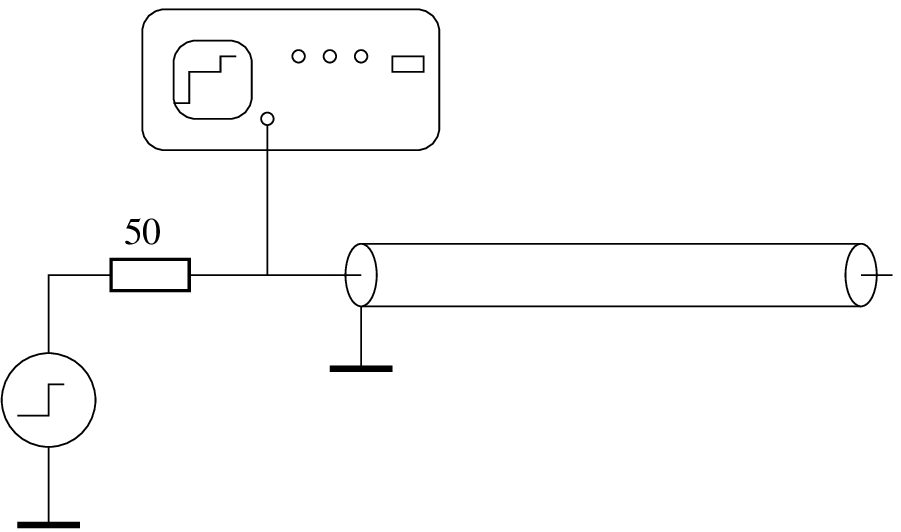}
   \caption{Time Domain Reflectometry setup}
   \label{fig:tdrsetup}
  \end{minipage}
  \begin{minipage}[c]{0.49\linewidth}
   \center\includegraphics[width=0.8\linewidth]{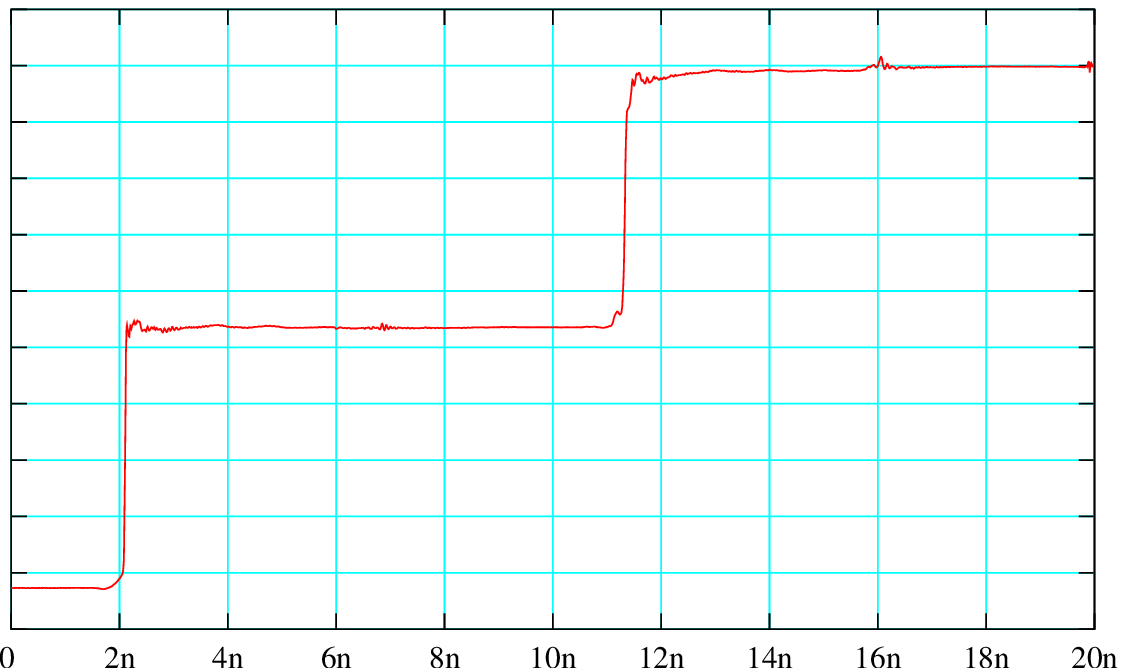}
   \caption{TDR plot of a 1 meter piece of coax}
   \label{fig:octdr}
  \end{minipage}
 \end{center}
\end{figure}

Figure~\ref{fig:tdrsetup} shows a simple example setup. A step signal is
launched into a 1~m piece of $50~\Omega$ coaxial cable. When the step hits
the open end, a positive reflection with the same size as the incident
step travels back towards the source.  Figure~\ref{fig:octdr} shows the
measured response, taken with a 50-year old Tektronix sampling scope. 
The reflection gets back just before the 12~ns mark.  The single-trip cable
delay is just under 5~ns, half of the distance between the two steps.
The T-junction connecting the transmission line to the oscilloscope must
be compact.                                                      

\begin{figure}[h]
 \center\includegraphics[width=.5\linewidth]{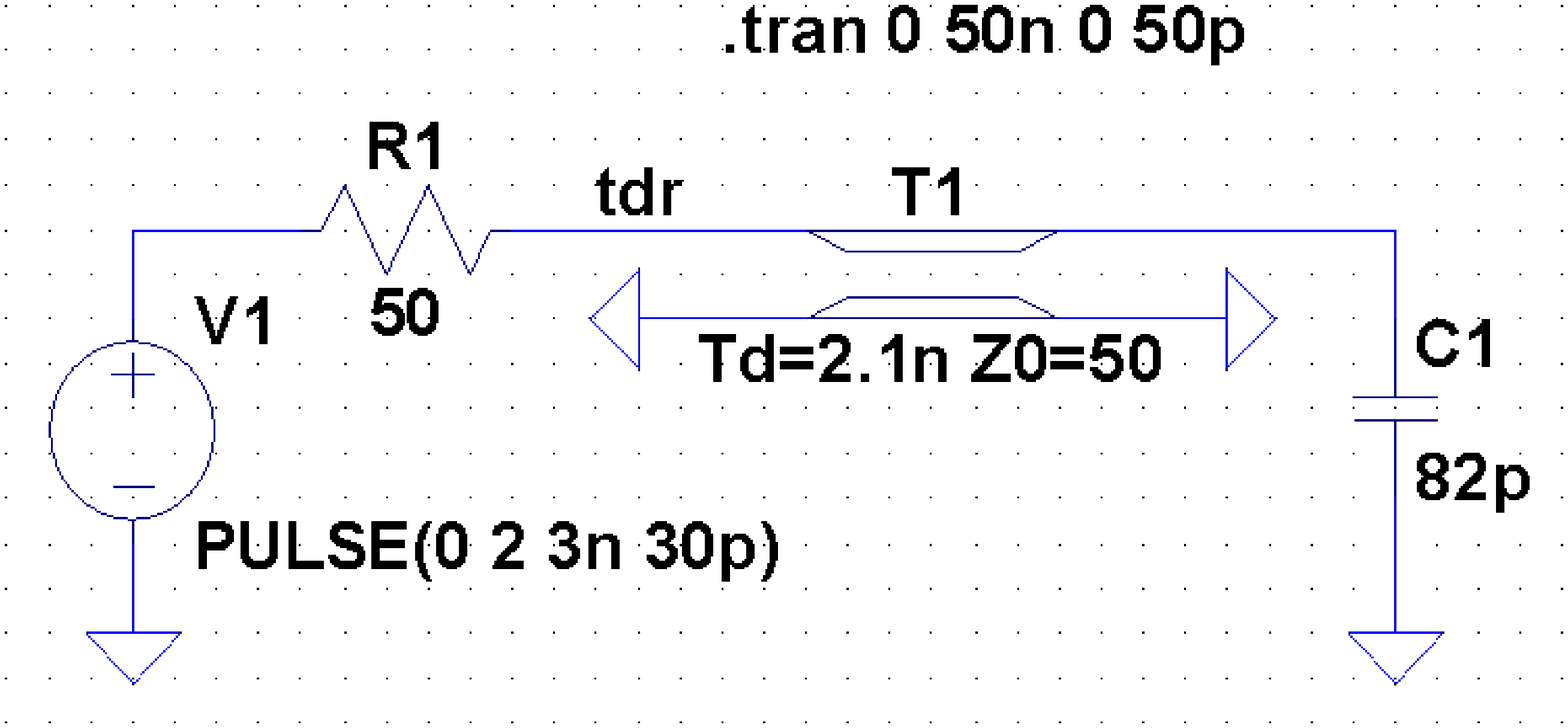}
 \caption{Simulation schematic of 2~ns of coax with an 82~pF capacitor at its end}
 \label{fig:captdr-sch}
\end{figure}

\begin{figure}[h]
 \begin{center}
  \begin{minipage}[c]{0.49\linewidth}
   \center\includegraphics[width=0.8\linewidth]{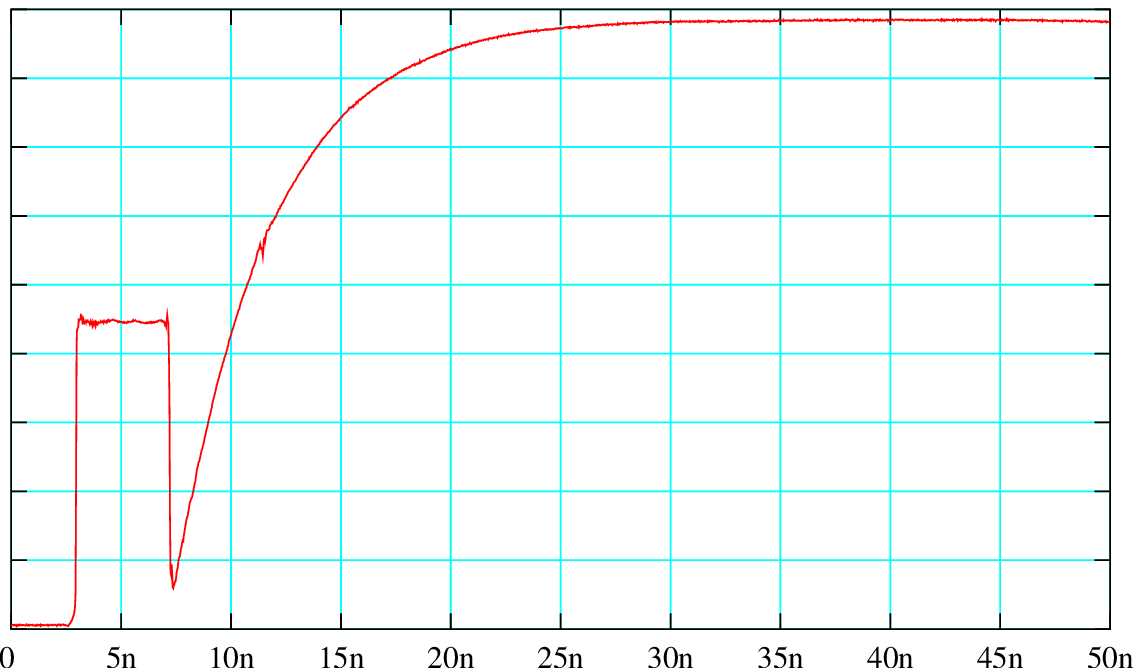}
   \caption{Measured reflection from a setup like in Fig.~\ref{fig:captdr-sch}}
   \label{fig:captdr}
  \end{minipage}
  \begin{minipage}[c]{0.49\linewidth}
   \center\includegraphics[width=0.8\linewidth]{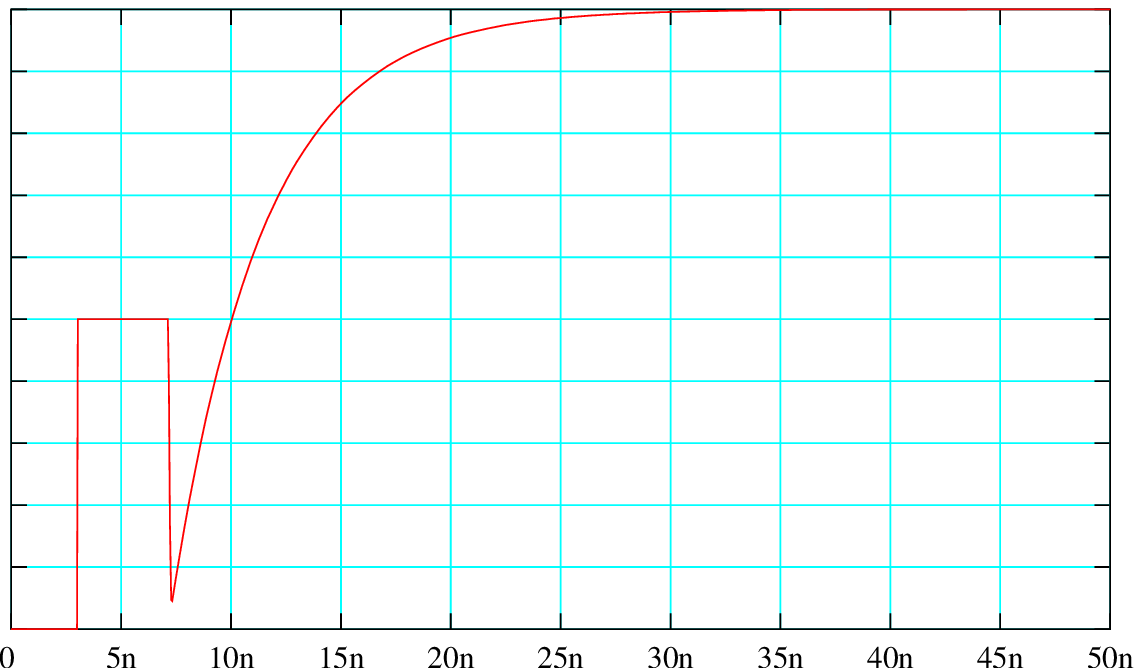}
   \caption{Simulated behaviour of Fig.~\ref{fig:captdr-sch}}
   \label{fig:captdr-sim}
  \end{minipage}
 \end{center}
\end{figure}

Another example, figure~\ref{fig:captdr-sch}, shows a piece of coax with
a capacitor at its end. Initially, the step generator sees just a piece
of $Z_0 = 50~\Omega$ coax. It has no way yet to know what's at the
far end.  When the initial step reaches the end, from the viewpoint of
the capacitor, this is just a step of twice the amplitude, applied from
a source with impedance $Z_0$.  The capacitor is charged according to
\begin{equation}
 \frac{U}{U_a} = 1-e^{\frac{-t}{Z_0 C}},
 \label{eq:captdr}
\end{equation}
where $U_a$ is the open-source magnitude of the applied step voltage.
This wave then travels back towards the generator to arrive there after
another cable delay.
Side by side comparisons of measured and simulated data can profitably
be used to build models of devices sounded by time-domain reflectometry,
as illustrated in figure~\ref{fig:captdr} and \ref{fig:captdr-sim},
showing the signals at the node marked
'tdr' in figure~\ref{fig:captdr-sch}.

A practical application of time-domain reflectometry
for us beam instrumentalists is the optimization
of a coaxial beam pick-up test bench. To test pick-ups or beam
transformers in the lab, we string a rod or wire through the pick-up
to which we apply test signals to simulate the passage of a beam.
We can use TDR techniques to match the ends of the rod and to reduce
any discontinuities. A smooth impedance guarantees a well-defined
current through the pick-up.
\begin{figure}[h]
 \center\includegraphics[width=0.8\linewidth]{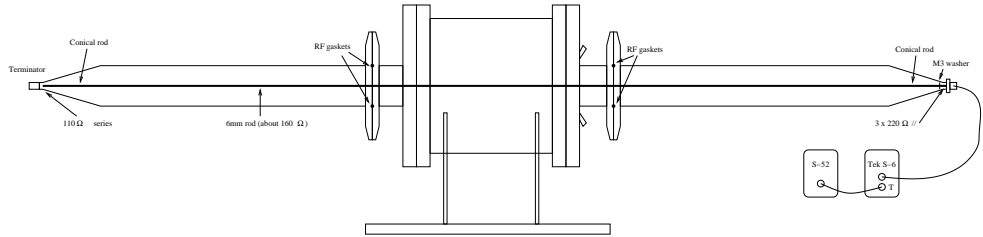}
 \caption{Test setup for a Wall Current Monitor}
 \label{fig:wcmtest}
\end{figure}
Reflections would clutter the response,
so it is good to carefully match both ends of the rod to minimize
them. 
With the 30~ps risetime of the test instrument, it's
pretty hard to reduce reflections to below 3\%.            
Precision geometry and RF-qualified components are needed to do
this right.
\begin{figure}[h]
 \begin{minipage}[b]{0.49\linewidth}
  \center\includegraphics[width=0.85\linewidth]{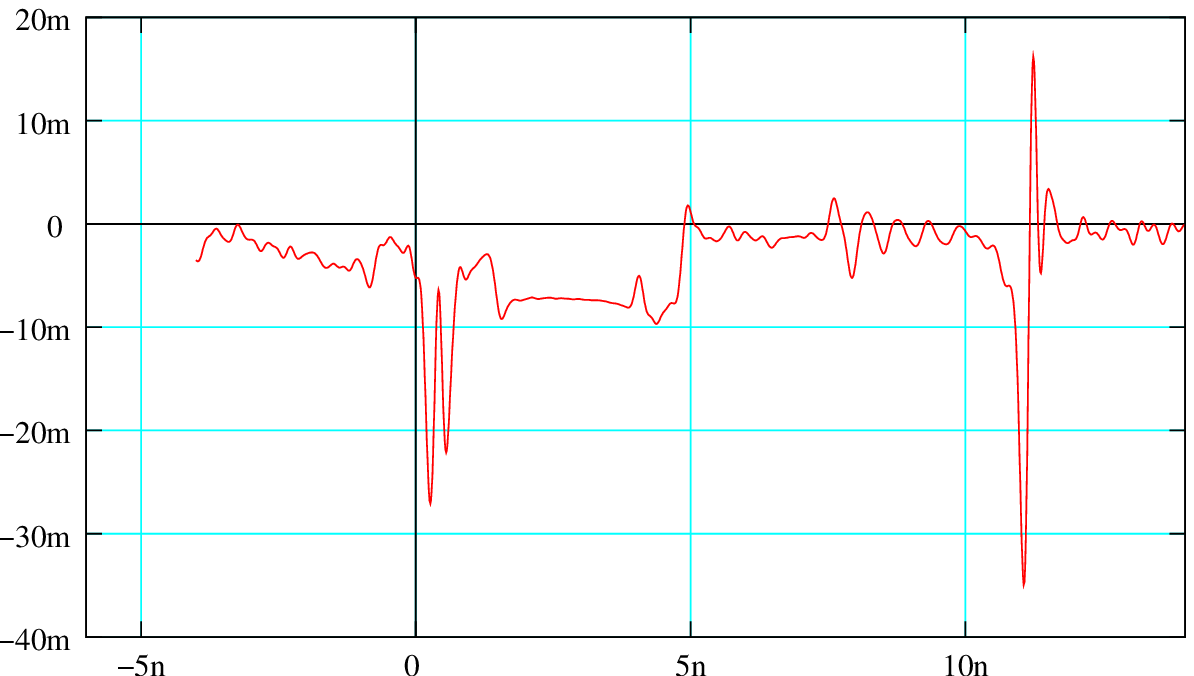}
  \caption{TDR plot of the rod through the WCM}
  \label{fig:rod2}
 \end{minipage}
 \begin{minipage}[b]{0.49\linewidth}
  \center\includegraphics[width=0.8\linewidth]{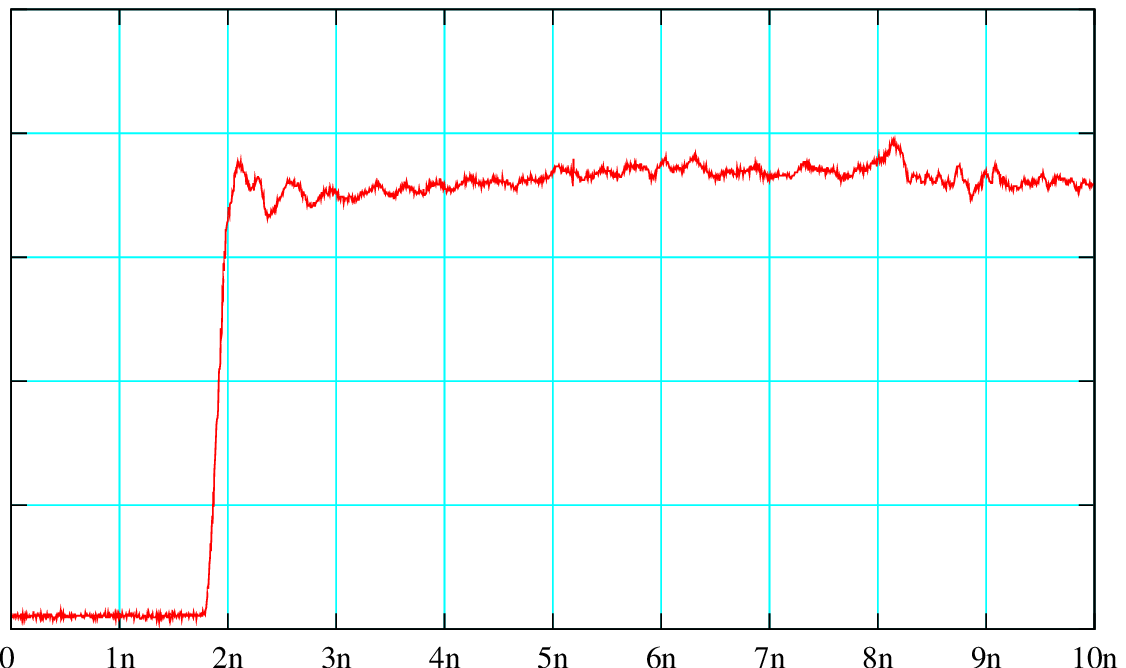}
  \caption{Step response of one of the ports of the WCM}
  \label{fig:port1}
 \end{minipage}
\end{figure}
Figure~\ref{fig:rod2} shows a typical TDR trace of this setup. The 
vertical scale was normalized to the incident step which is off
scale to the left of the plot. We recognize
the residual discontinuities of the rod ends (at 0 and 11~ns), the impedance
of the gap inside the Wall Current Monitor (at 5~ns) and we can identify
the gaps between the vacuum flanges at each end of the WCM (at 4 and
8~ns). 
Figure~\ref{fig:port1} shows the response of one of the output ports
of the WCM. The risetime is about 70~ps.

Note that a Fourier transform of differentiated time-domain reflection
data will yield the frequency-domain reflection coefficient, from
which the complex impedance can be found via Eq.~\ref{eq:bilinear}).
The reverse, applying an inverse Fourier transform to frequency-domain
reflection data from an RF network analyzer, followed by integration
over time will yield a TDR plot. Convolution in the frequency domain
instead of integration/differentiation in the time domain will also work.
Note that a TDR plot derived from frequency-domain data would show just
the reflection, without the incident signal, because of the directional
coupler in the network analyzer.

\subsection{Transmission line transformers}
 
Transmission line transformers are useful in analog signal processing
as matching circuits to change voltage, current and impedance levels,
as baluns and inverters, combiners, splitters, hybrids and directional
couplers.  The accessory fact that they are rad-hard is a useful feature
for their use in beam instrumentation.

Ordinary transformers are bandwidth-limited by magnetic core losses, by
leakage inductance and by parasitic capacitance.  The useful frequency
range can be extended considerably by arranging the windings to take
advantage of transmission line coupling.
In the context of RF transformers, it is customary to specify the
impedance ratio rather than the turns ratio. The impedance ratio is
simply the square of the turns ratio.

\begin{figure}[h]
 \begin{minipage}[c]{0.49\linewidth}
  \center\includegraphics[width=0.6\linewidth]{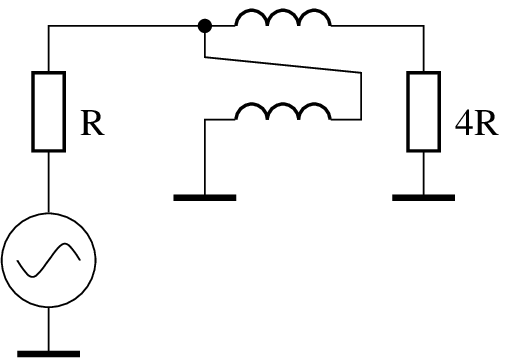}           
 \end{minipage}
 \begin{minipage}[c]{0.49\linewidth}
  \center\includegraphics[width=0.6\linewidth]{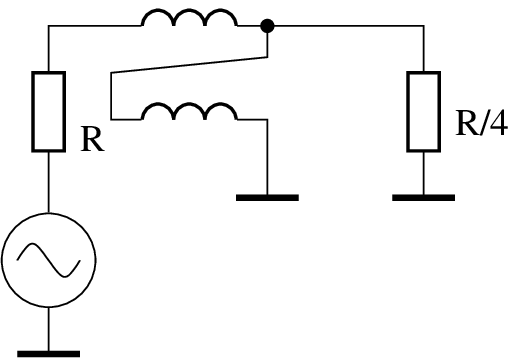}
 \end{minipage}
 \caption{Ruthroff-type transmission line transformers}
 \label{fig:ruthroff-wire}
\end{figure}

\begin{figure}[h]
 \begin{minipage}[c]{0.45\linewidth}
  \center\includegraphics[width=0.6\linewidth]{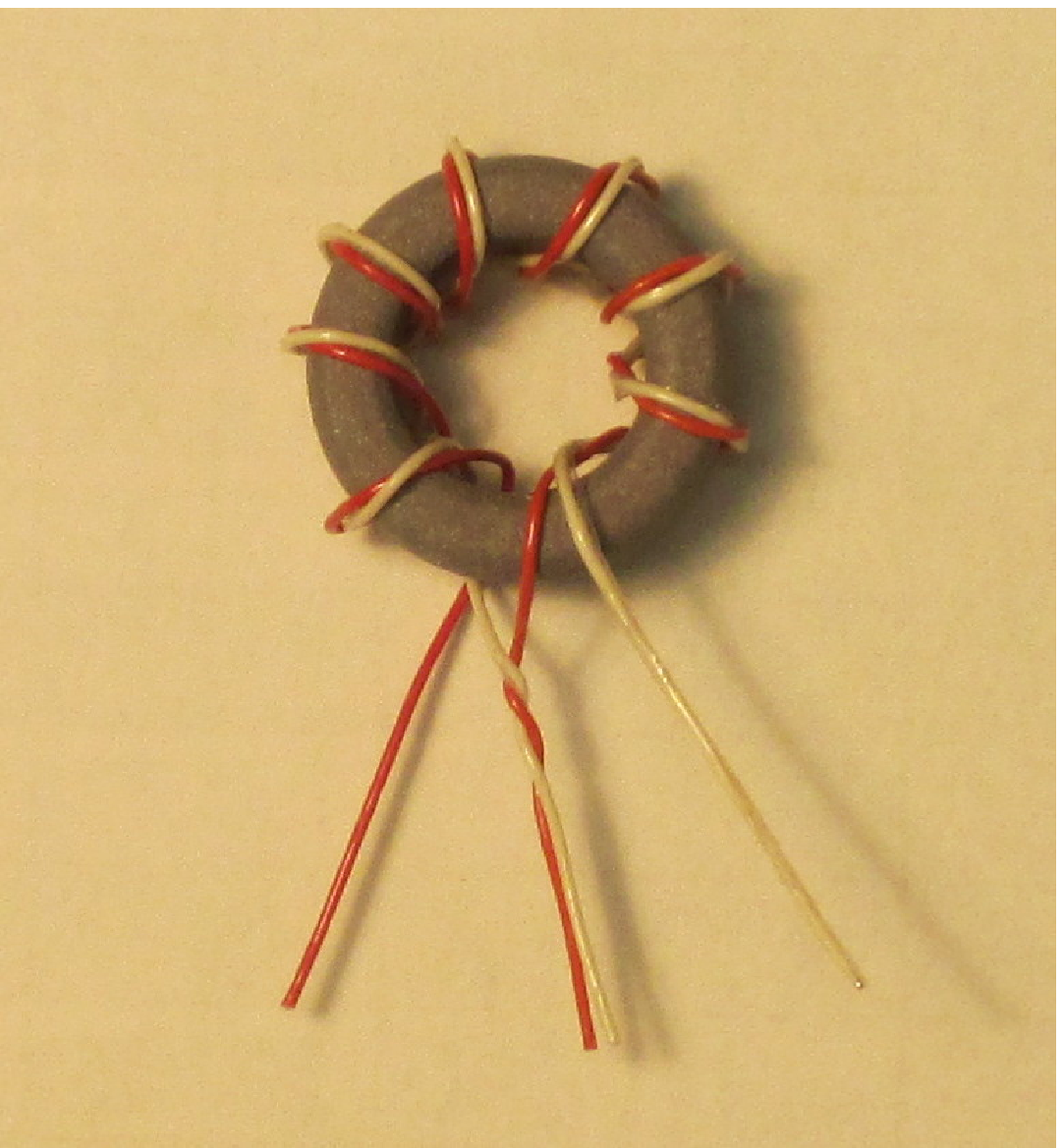}
  \caption{Construction of a wire-wound 1-4 Ruthroff transformer}
  \label{fig:Ruthroff1-4-picture}
 \end{minipage}
 \hfill
 \begin{minipage}[c]{0.49\linewidth}
  \center\includegraphics[width=1\linewidth]{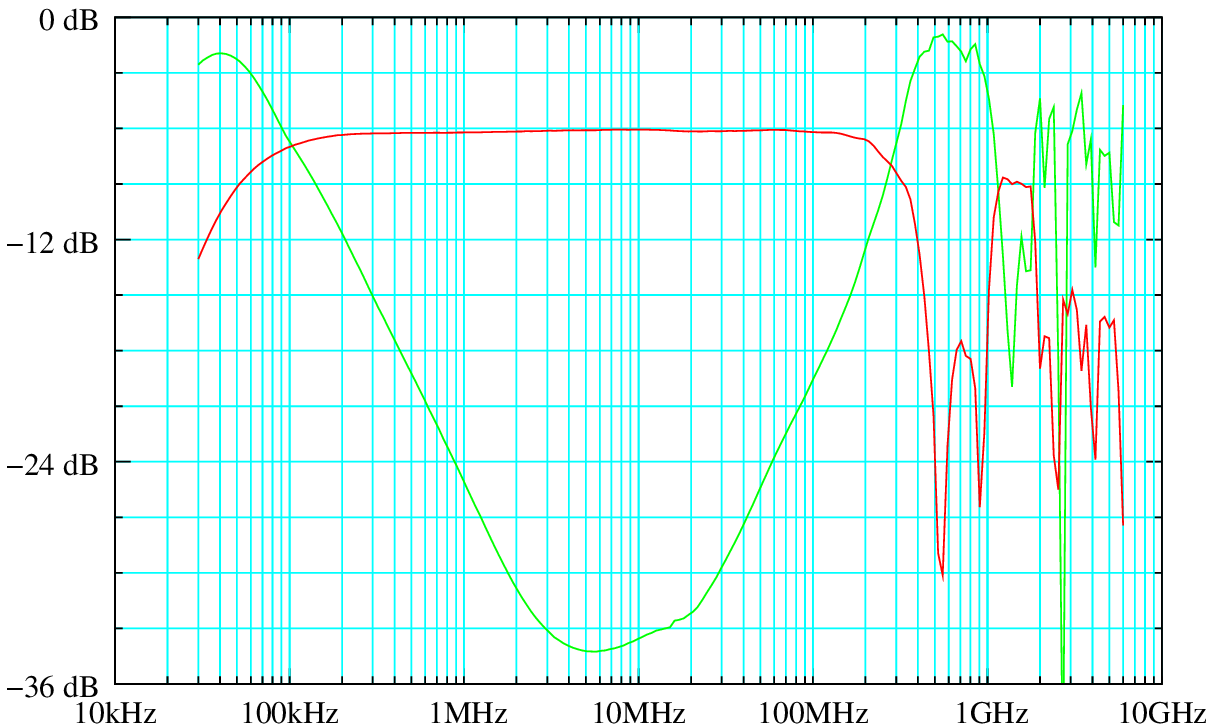}
  \caption{Frequency response of a wire-wound 1-4 Ruthroff transformer}
  \label{fig:Ruthroff1-4}
 \end{minipage}
\end{figure}

A basic example is the 1:4 Ruthroff transformer (Fig.~\ref{fig:ruthroff-wire},
\ref{fig:Ruthroff1-4-picture})~\cite{bib:ruthroff1959}. The low-frequency
equivalent circuit is just an autotransformer with a tap in the middle.
The windings are laid down as a closely-spaced pair of wires. This ensures
that transmission line coupling takes over at frequencies where magnetic
coupling through the flux in the core fails, thus extending the useful
bandwidth. About three decades of bandwidth are achievable.
These transformers have a null in the frequency response where
the wavelength in the transmission line formed by the wire pair is
$\lambda/2$, because the delayed signal from the output end of the
transformer is fed back into the input. This implies that the wires
must be kept as short as possible to preserve the high frequency
performance.

Figure~\ref{fig:Ruthroff1-4} (red curve) shows the measured frequency
response of the transformer in figure~\ref{fig:ruthroff-wire}, left.  On
the input side, a $50~\Omega$ source drives the circuit.  The transformer
increases the the voltage at its input port by a factor of two, +6~dB.
On the output side, the load must be $200~\Omega$, composed here of a
$150~\Omega$ resistor in series with the $50~\Omega$ input resistance
of the measurement instrument, an RF Network Analyzer. This forms a
resistive divider of 1/4, or -12~dB. Thus the level of the flat portion
of the response as seen by the instrument is -6~dB.
The lower cut-off frequency is where the inductive reactance of the
winding drops below $25~\Omega$. 
The upper cut-off frequency occurs where the length of the winding
reaches $1/4~\lambda$ and there is a null at twice that frequency.
The transformer incurs a little loss, usually well below 1~dB in the
flat portion of the response.
The green curve is the reflection
coefficient in decibels, a measure of how closely the transformer's
input impedance matches 50~$\Omega$. (See also Eq.~(\ref{eq:bilinear})).

It is possible to obtain other transformation ratios by adding or
removing one or two turns from one of the windings, or by using
three or more wires instead of just two. Even though this stretches
the concept of a transmission line a bit,
results are often quite acceptable~\cite{bib:sevick}.

\begin{figure}[h]
 \begin{minipage}[c]{0.49\linewidth}
  \center\includegraphics[width=0.8\linewidth]{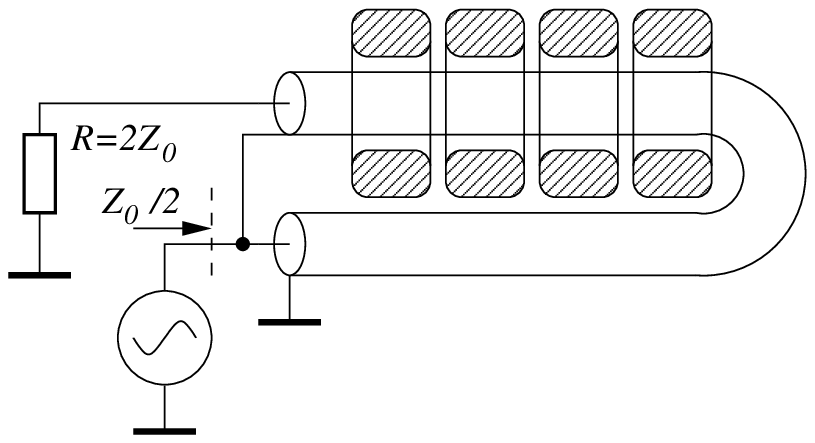}           
 \end{minipage}
 \begin{minipage}[c]{0.49\linewidth}
  \center\includegraphics[width=0.8\linewidth]{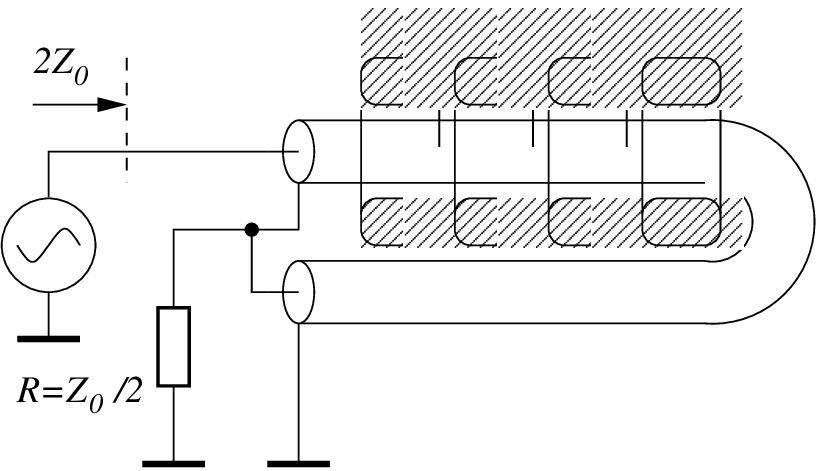}
 \end{minipage}
 \caption{Ruthroff-type impedance transformation with coax}
 \label{fig:ruthroff-coax}
\end{figure}

It is also possible to use coax cable for the windings
(Fig.~\ref{fig:ruthroff-coax}). The centre conductor replaces one wire
and the screen conductor replaces the other.
It may be more practical to thread the coax
straight through multiple cores rather than to try to wind it on a
single core.
The coax impedance should ideally be the geometrical mean of input
and output impedances. It is advisable to select the screen for
the conductor with the smallest signal potential, to minimize the
effects of parasitic capacitances.
The advantage of coax is the possibility to obtain lower leakage 
inductance, allowing higher frequency operation. 
Otherwise, the same reasoning as for the
wire-wound transformers applies.

The interesting thing about transmission line transformers is that their
performance at high frequencies depends but little on the properties of
the magnetic cores. The cores serve only to impede common-mode current.
The core and windings are dimensioned so that the common-mode inductive
reactance at the lower end of the design bandwidth is greater than the
combined port impedances. In addition, the core size should be such that
it does not saturate at the rated power level and the lowest frequency.
The usual core choices are toroids of high-permeability ferrite, or
amorphous or nano-crystalline magnetic alloys.

\subsubsection{Balun transformers}

\begin{figure}[h]
 \begin{minipage}[b]{0.49\linewidth}
  \center\includegraphics[width=0.6\linewidth]{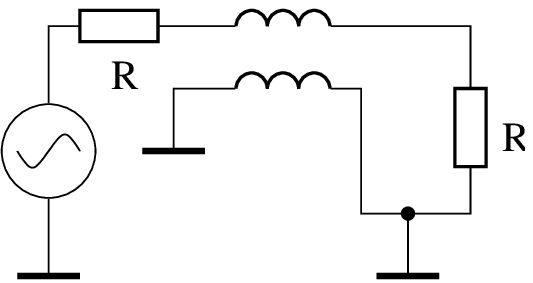}
 \end{minipage}
 \begin{minipage}[b]{0.49\linewidth}
  \center\includegraphics[width=0.6\linewidth]{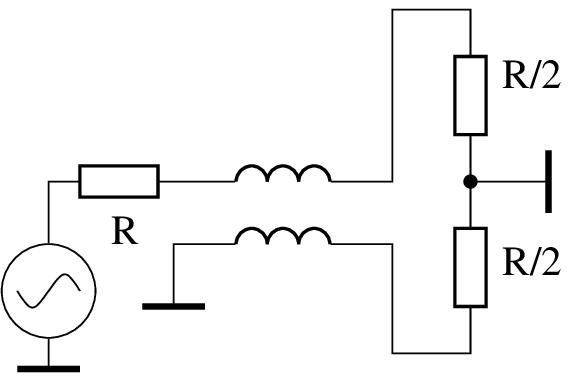}
 \end{minipage}
 \begin{minipage}[b]{0.50\linewidth}
  \center\includegraphics[width=0.6\linewidth]{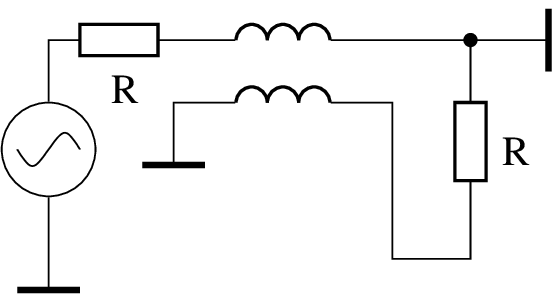}
 \end{minipage}
 \begin{minipage}[b]{0.50\linewidth}
  \center\includegraphics[width=0.6\linewidth]{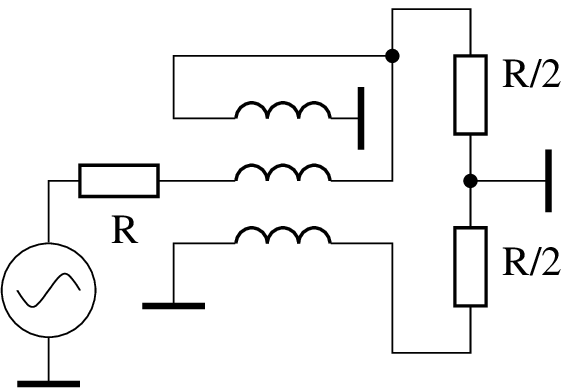}
 \end{minipage}
 \caption{Some balun variants}
 \label{fig:baluns}
\end{figure}

The word \emph{balun} is a contraction of
\emph{bal}anced-to-\emph{un}balanced.  The general topology is a
transformer turned on its side (Fig.~\ref{fig:baluns}).  Both windings
are wound together on a single core, in a similar way as in figure~
\ref{fig:Ruthroff1-4-picture}.
The idea is
that the left and right sides of the transformer are separated by the common-mode
inductance of the windings, lending almost complete freedom in choosing
the grounding points on either side. This works for frequencies where
the common-mode impedance becomes
greater than the source and load impedances.

In the examples shown in figure~\ref{fig:baluns}, the input source
is always single-ended and referenced to ground, but this is entirely
optional.  The top left diagram is the trivial case, with the bottom
side of the load resistance R also connected to ground.  The load
sees an untransformed signal, albeit delayed by the length of the
transmission line. At the bottom left, the load resistance is grounded
at the other end.  The balun now acts as an inverter.  In the top right
diagram, the centre of the load resistance is connected to ground. The
load now gets driven symmetrically. This is the application from which
this class of transformers derives its name.  A minor issue is that
there is still some current flowing through the common mode impedance
of the transformer, which also flows through the top half of the load,
but not through the bottom half. As a result, the symmetry is a little
off, worse at low frequencies.  The extra winding in the bottom right
diagram fixes that by providing an alternative path for that current.

\begin{figure}[h]
 \begin{minipage}[c]{0.49\linewidth}
  \center\includegraphics[width=0.8\linewidth]{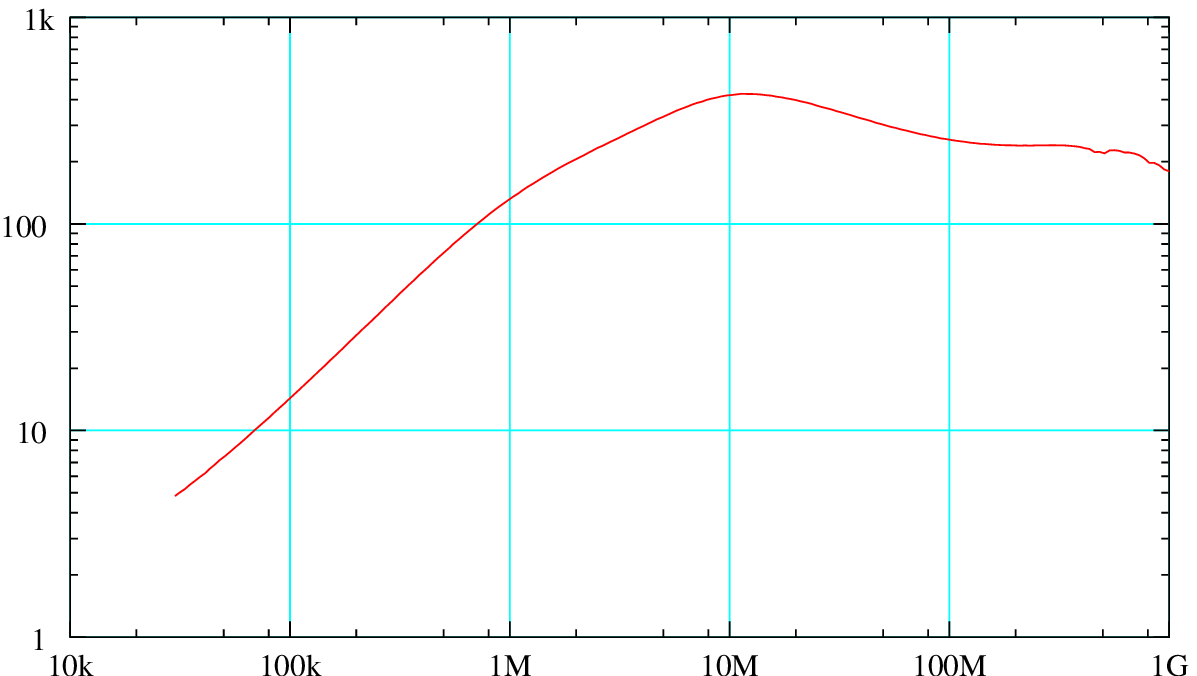}
  \caption{Impedance vs. frequency of a 6-turn coil on a small high-permeability ferrite toroid core}
  \label{fig:CMZ} 
 \end{minipage}
 \begin{minipage}[c]{0.49\linewidth}
  \center\includegraphics[width=0.7\linewidth]{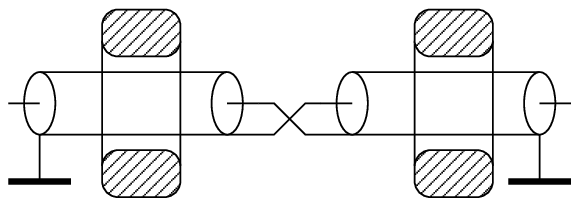}
  \caption{Inverting transformer schematic}
  \label{fig:xform-cir}
 \end{minipage}
\end{figure}

Again, the necessary common-mode inductance may be obtained by winding the
conductors on a high-permeability toroid cores.
Here too, the high-frequency properties of
the cores hardly matter.  Even though core losses cause the common mode
impedance to become resistive --rather than inductive-- above a few megahertz,
it will usually remain comfortably higher than the winding's differential
mode impedance, which is sufficient (Fig.~\ref{fig:CMZ}).

\begin{figure}[h]
 \begin{minipage}[c]{0.49\linewidth}
  \center\includegraphics[width=0.8\linewidth]{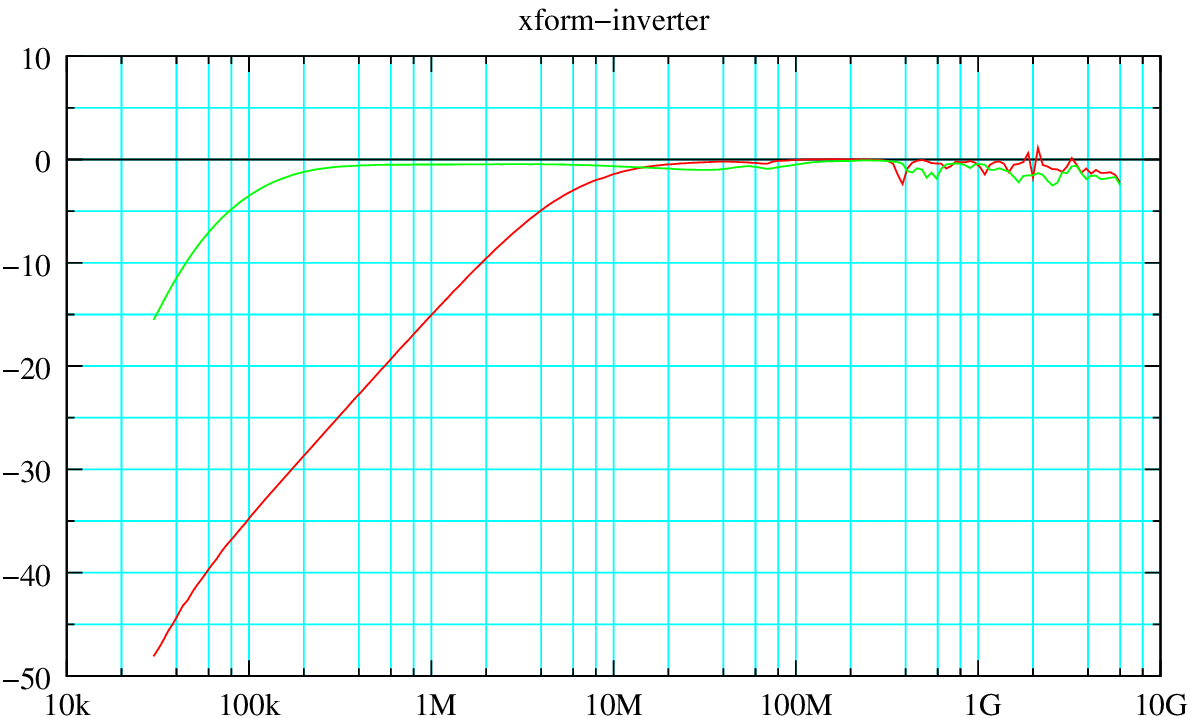}
  \caption{Simple transmission line inverting transformer frequency response}
  \label{fig:xform-inverter}
 \end{minipage}
 \begin{minipage}[c]{0.49\linewidth}
  \center\includegraphics[width=0.8\linewidth]{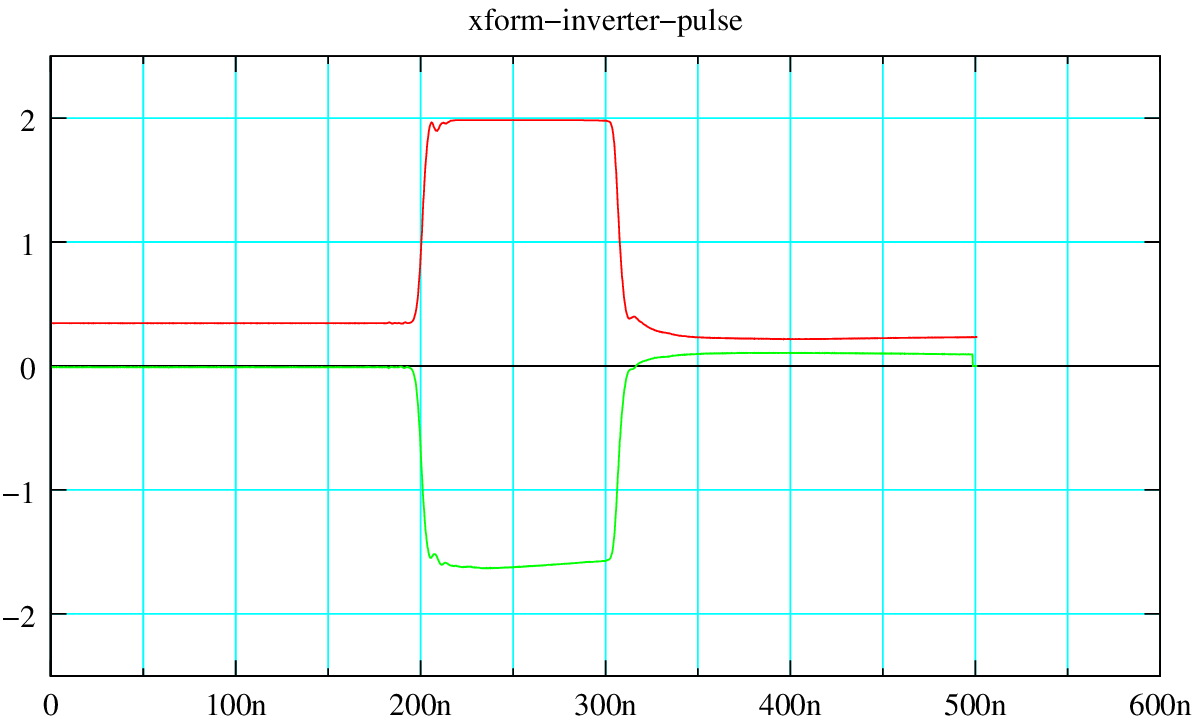}
  \caption{Simple transmission line inverting transformer pulse response}
  \label{fig:xform-inverter-pulse}
 \end{minipage}
\end{figure}

Again, coax cable can be used instead of wire.
A simple example implements the inverting transformer of 
figure~\ref{fig:baluns}, bottom left:
Cross the shield and core conductors near the middle of an
ordinary piece of $50~\Omega$ coax 60~cm long
and you'd have an excellent
inverting transformer (Fig.~\ref{fig:xform-cir}).
A 60~cm length of $50~\Omega$ coax
--any electronics lab has lots of those-- has a common mode inductance
of about half a microhenry, the impedance of which exceeds the
characteristic impedance of the coax above about 8~MHz.  
So the useful frequency range starts at 8~MHz and reaches up to over 6~GHz
(Fig.~\ref{fig:xform-inverter}, red curve).

But it gets better: By slipping a handful of ferrite cores over the
cable, it's easy to raise the common mode inductance to  several tens of
microhenries. The transformer then also works at much lower frequencies,
like the green curve in figure~\ref{fig:xform-inverter}.  With little
effort, we have made an inverting transformer with a bandwidth of 100~kHz
to over 6~GHz. Pretty impressive for such a simple thing! Since above
8~Mhz the EM fields are entirely confined inside the coax, the ferrite
properties are largely irrelevant there.  The core has to 'work'
only from about 100~kHz to 8~MHz.

\subsubsection{Equal delay transformers}

\begin{figure}[h]
 \begin{minipage}[b]{0.5\linewidth}
  \center\includegraphics[width=0.9\linewidth]{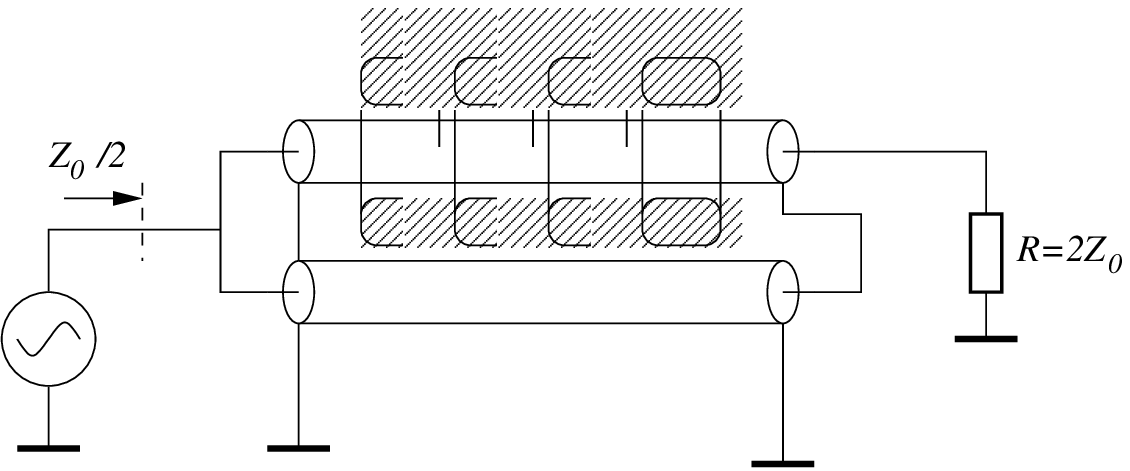}
  \caption{A Guanella 1:4 transformer}
  \label{fig:Guanella1} 
 \end{minipage}
 \begin{minipage}[b]{0.5\linewidth}
  \center\includegraphics[width=0.9\linewidth]{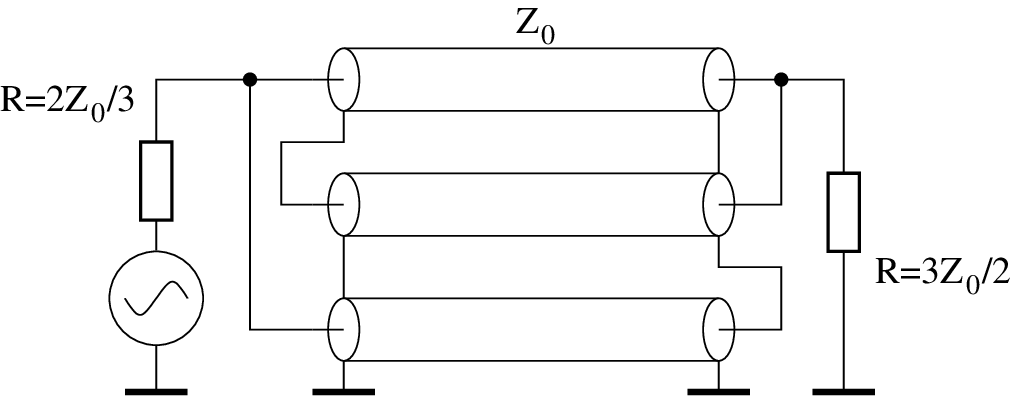}
  \caption{A Guanella 1:2.25 transformer}
  \label{fig:ravor2} 
 \end{minipage}
\end{figure}

Baluns can be combined to create \emph{equal delay} or \emph{Guanella}
transformers \cite{bib:guanella1944}.  Here, the idea is to effect
impedance transformations by connecting transmission line baluns
in series(parallel) on one side, and in parallel(series) on the
other. If the transmission lines have equal lengths, signals add in
phase at the output and the frequency response is flat.
Figure~\ref{fig:Guanella1} shows such a transformer with a 1:4 ratio.
The high-end frequency is limited only by the inevitable inductance of the
interconnections, by the capacitance to the upper transmission
line, which has an RF voltage on its screen, and by length mismatches
between the lines.
Implemented with $50~\Omega$ coax, it would have a $25~\Omega$ impedance
at the left side and a $100~\Omega$ impedance at the right. It is not
unreasonable to expect the bandwidth to stretch from a few tens of
kiloherz up to several gigaherz.

The impedance ratio of Guanella transformers is not limited to only the
squares of the integers.  In principle, any rational transformation ratio
can be obtained by various combinations of parallel and series connections
of transmission lines, see, for example, figure~\ref{fig:ravor2}. (The
necessary ferrite cores for the top two lines are not shown. The bottom
line needs none because there is no common mode voltage across the
ends.) Many combinations are detailed in \cite{bib:mcclure1994}.  In
practice the range is limited by parasitic inductances and capacitances,
and by the difficulty of joining many transmission lines in a compact way.

\subsubsection{Combiners, splitters and hybrids}
Another class of transmission line transformers combines
the signals from two independent sources together on a single output,
while preventing any power flow between the inputs.
\begin{figure}[h]
 \begin{minipage}[b]{0.5\linewidth}
  \center\includegraphics[width=0.7\linewidth]{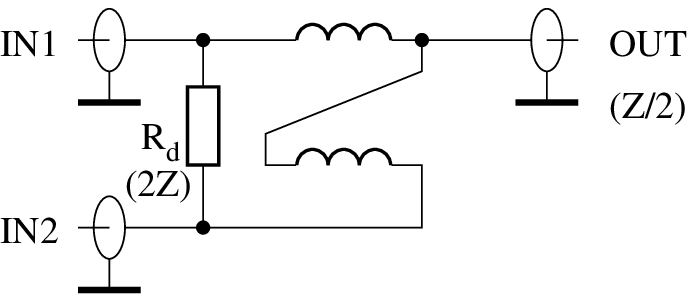}
  \caption{A wire-wound in-phase combiner}
  \label{fig:combiner1}
 \end{minipage}
 \begin{minipage}[b]{0.5\linewidth}
  \center\includegraphics[width=0.6\linewidth]{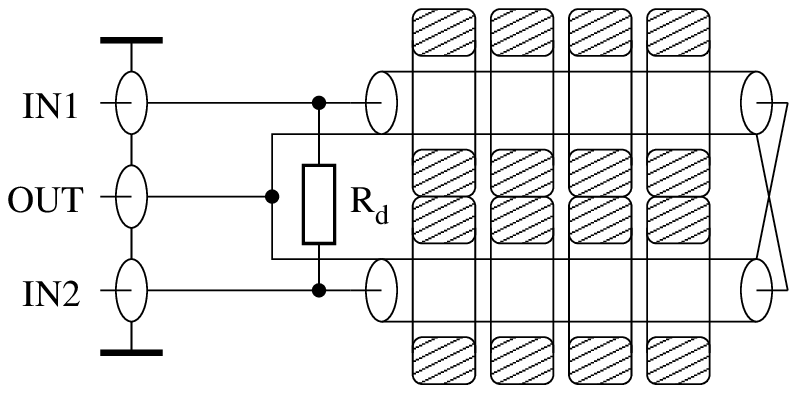}
  \caption{An in-phase combiner with coax}
  \label{fig:combiner2}
 \end{minipage}
\end{figure}
Figure~\ref{fig:combiner1} shows a wire-wound version. The two input
signals are added together and appear at the output at half the original
source impedance. Resistor $R_d$ is necessary to provide isolation between
the inputs and absorbs the \emph{difference} between the inputs. Its
value is \emph{twice} the original source impedance, because it sees
the inputs in series, whereas the sum output sees the inputs in parallel
instead.  Figure~\ref{fig:combiner2} shows a possible implementation of
the same thing using coaxial cable. Using either transformer in reverse,
exchanging inputs and outputs, they would operate as splitters.

\begin{figure}[h]
 \begin{minipage}[b]{0.5\linewidth}
  \center\includegraphics[width=0.7\linewidth]{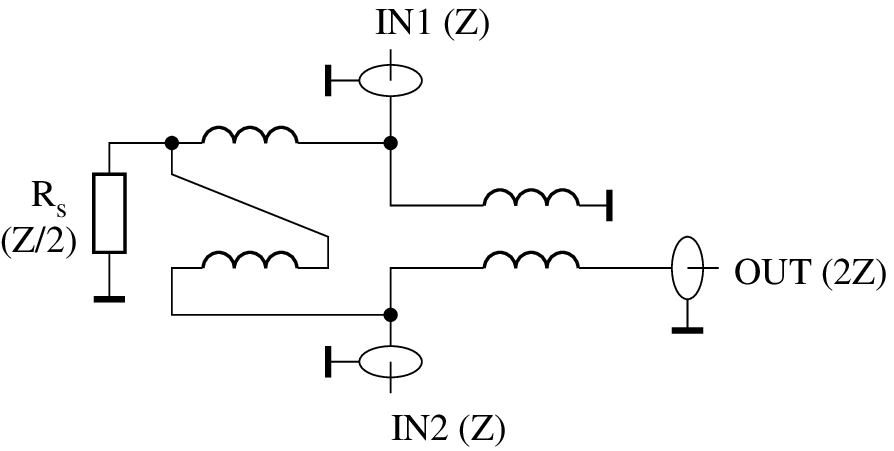}
  \caption{A wire-wound 180$^\circ$ combiner}
  \label{fig:combiner3}
 \end{minipage}
 \begin{minipage}[b]{0.5\linewidth}
  \center\includegraphics[width=0.5\linewidth]{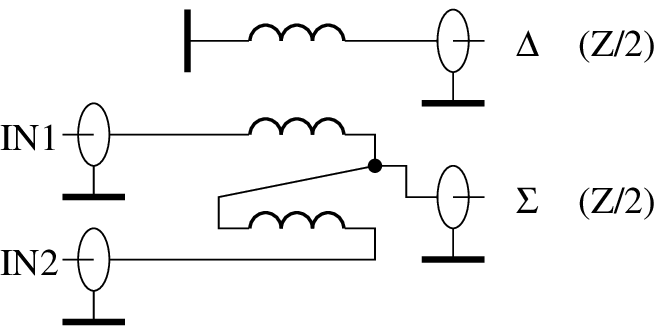}
  \caption{A wire-wound hybrid}
  \label{fig:hybrid}
 \end{minipage}
\end{figure}
A minor change turns the circuit of figure~\ref{fig:combiner1} into
a 180$^\circ$ combiner, subtracting the inputs rather than adding them (Fig~\ref{fig:combiner3}).
Instead of dumping the difference into a resistor,
we use a balun to transform this balanced signal to a
single-ended output. We now dump the sum signal into a
resistor. 

Figure ~\ref{fig:hybrid} shows a different way to extract a single-ended
difference. In addition, this transformer also puts out the sum at the
same time, so it's now a hybrid.  The three wires are wound on a
single core. If the number of turns is the same for all three windings,
the difference signal will also have an impedance of $Z_0/2$, like
the sum.
Such hybrids are often used in beam position monitor signal
processing. 

\begin{figure}[h]
 \center\includegraphics[width=0.45\linewidth]{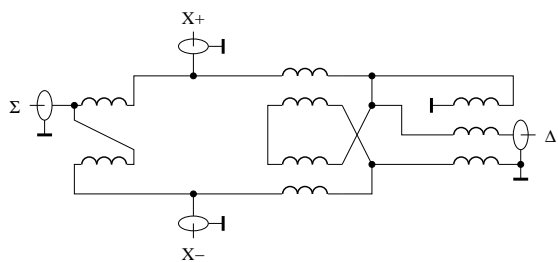}
 \caption{A wide-band hybrid}
 \label{fig:4Dhybrid-sch}
\end{figure}

\begin{figure}[h]
 \begin{minipage}[c]{0.55\linewidth}
  \center\includegraphics[width=0.9\linewidth]{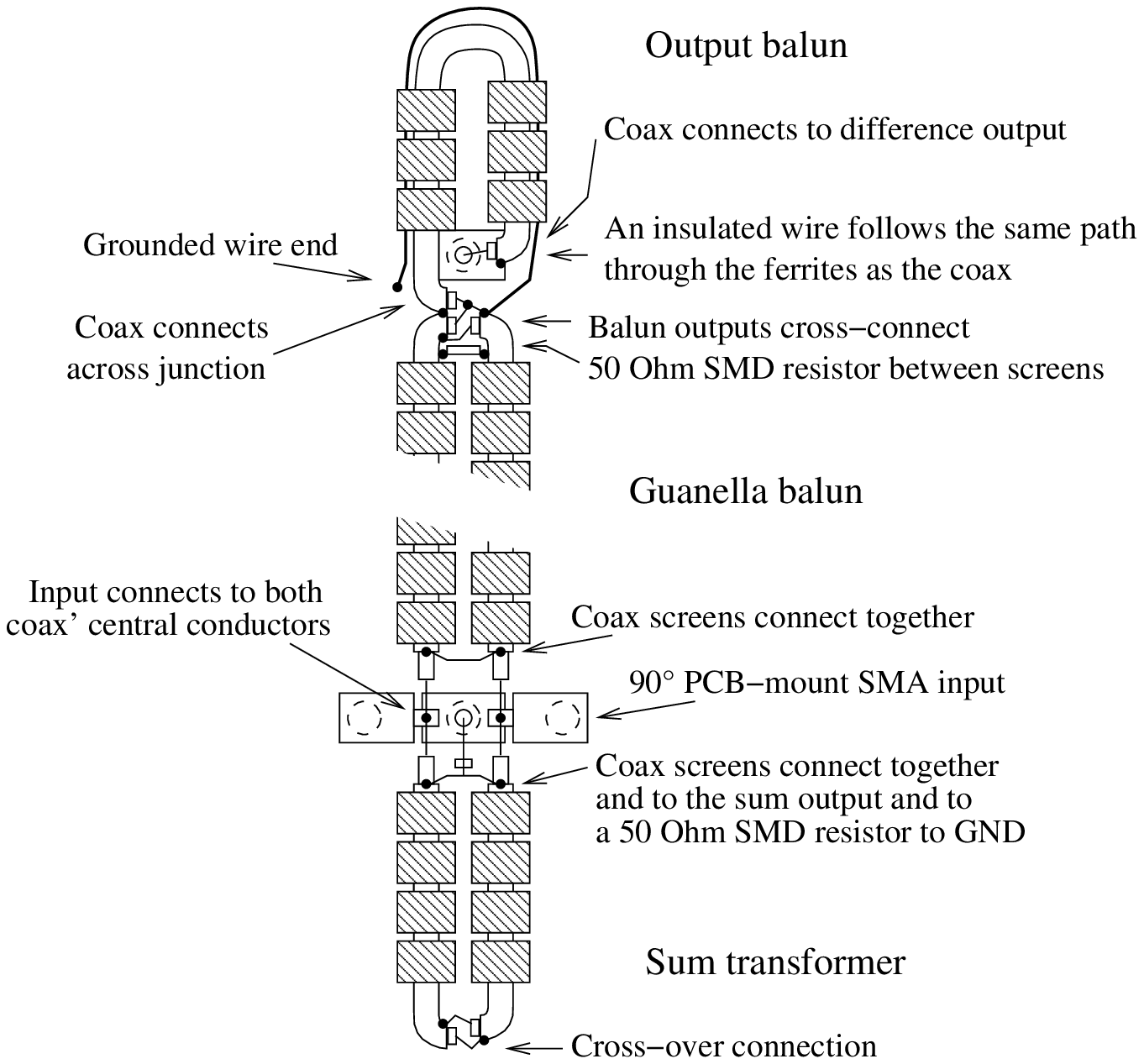}
  \caption{Practical implementation of the wide-band hybrid}
  \label{fig:4Dhybrid}
 \end{minipage}
 \begin{minipage}[c]{0.45\linewidth}
  \center\includegraphics[width=1\linewidth]{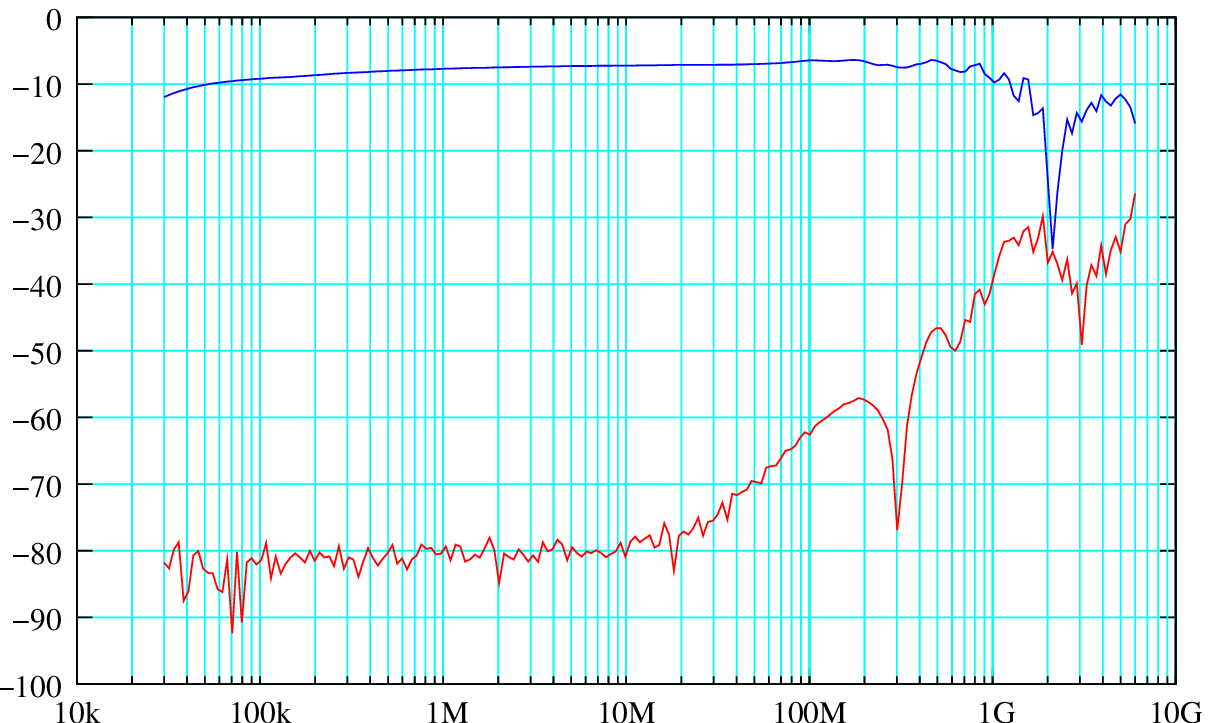}
  \caption{Frequency response of wide-band hybrid, top:$\Sigma$, 
           bottom:$\Delta$, with identical inputs.}
  \label{fig:YoverS}
 \end{minipage}
\end{figure}

Combining several of the transformers described above, it's
possible to construct hybrids with impressive bandwidths.
Figure~\ref{fig:4Dhybrid-sch} shows the simplified schematic of
a hybrid covering a bandwidth from 6~kHz to 600~MHz. It uses a
Ruthroff-type transformer for the sum output and a cascade of a
Guanella balun and a three-wire balun to construct the difference
output. 
Figure~\ref{fig:4Dhybrid} shows the practical construction, using thin
coaxial cable and one piece of wire. This hybrid is used for several
wide-band beam position monitors in CERN's PS complex \cite{bib:4Dhybrid}.
Figure~\ref{fig:YoverS} shows the measured frequency response of the
hybrid. For this application, the useful bandwidth is limited at the
high end, not so much by signal drop-off, as by the leakage of the sum
signal into the difference output.

\section{Filters}
Passive LC filters are indispensable between a transducer and
the first stage of active electronics.  They protect against out-of-band
signals and reduce the bandwidth and dynamic range demanded of amplifiers
and ADCs.  They improve the signal to noise ratio by rejecting
out-of-band noise.  Passive filters are also used as reconstruction
filters after a DAC, PWM or class-D output stage.

Many passive filters were developed and categorized on some useful
properties in the 1940's. Pre-computed tables of normalized element
values exist for filters that are optimized for clean impulse response
or steep frequency response, with variants trading some parameters against
others~\cite{bib:Zverev}. Trade-offs can be made against properties such
as steepness, passband ripple, ultimate attenuation, phase linearity
and complexity.
The calculation of element values ab initio is beyond the scope of
this text~\cite{bib:Saal1958}.
%
%

Passive filters are designed to provide the intended response with a
specified source and load impedance.  It's important to realize that
stop band energy is reflected back to the source, which must be able
to absorb it.  We shall see a few examples of filter designs based on
tables of normalized element values. Many commercial software tools 
exist to assist with filter design.

\begin{figure}[h]
 \begin{minipage}[c]{0.45\linewidth}
  \includegraphics[width=1\linewidth]{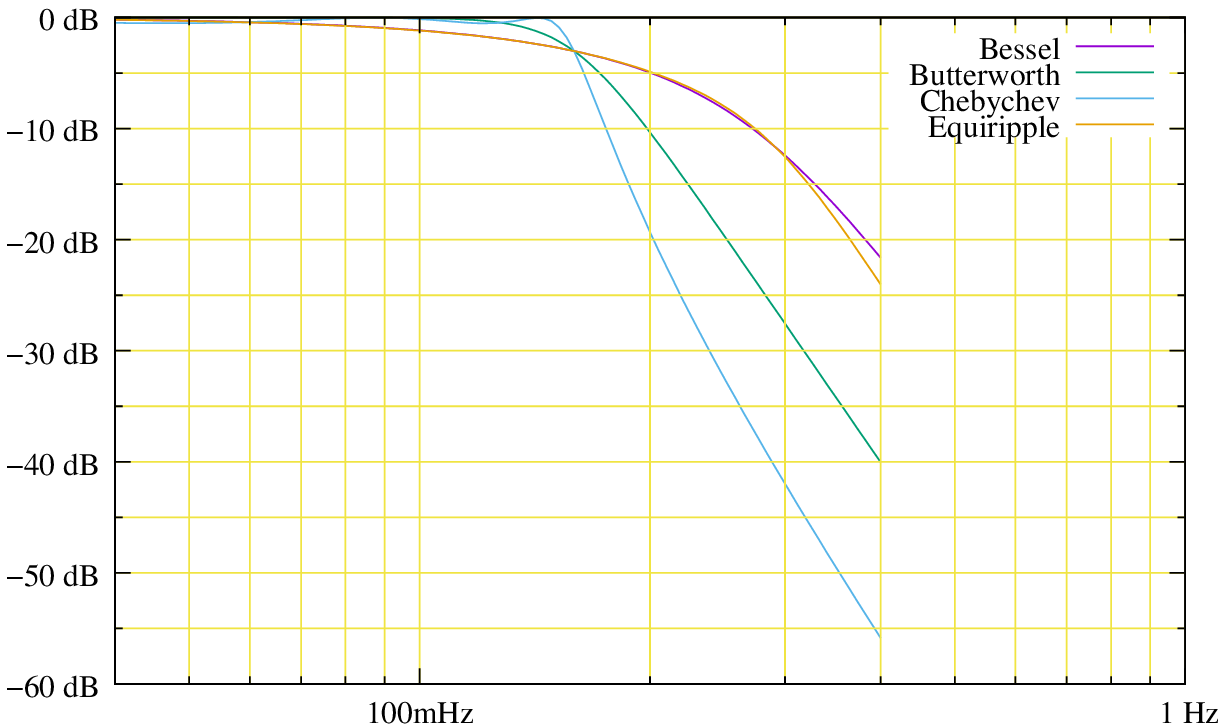}
  \caption{Normalized frequency response of some standard filters}
  \label{fig:multifilter}
 \end{minipage}
 \hfill
 \begin{minipage}[c]{0.45\linewidth}
  \includegraphics[width=1\linewidth]{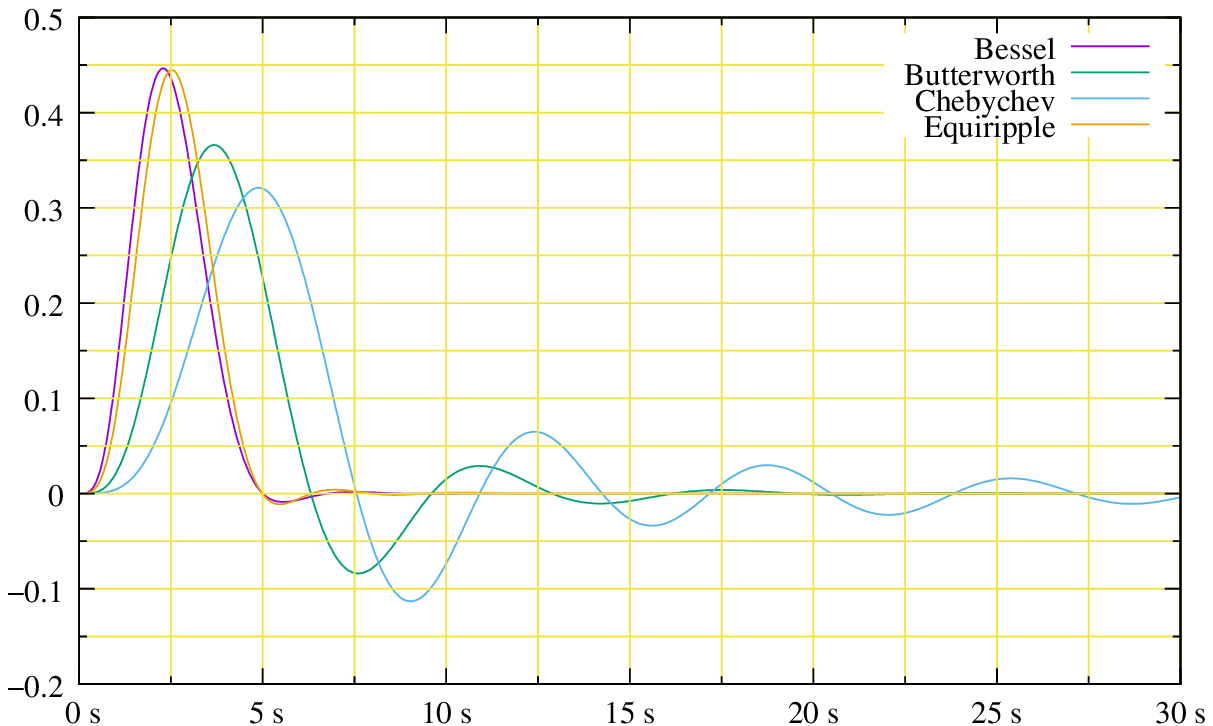}
  \caption{Normalized impulse response of some standard filters}
  \label{fig:multifilterimp}
 \end{minipage}
\end{figure}

If the application requires good pulse shape fidelity, we use filters that
approximate linear-phase behaviour --constant group delay--, such as the
Gaussian, Bessel or equi-ripple phase error filters. Such filters have
gently sloping frequency responses.  If the application calls for a steep
drop-off, Chebyshev or Cauer filters might be chosen, but the phase
response will be poor and the impulse response will ring.
The Butterworth filter is in between, falling off faster than the Bessel
but having better impulse response than a Chebyshev. See the curves in
figures \ref{fig:multifilter} and \ref{fig:multifilterimp}.

\begin{figure}[h]
\center\includegraphics[width=.7\linewidth]{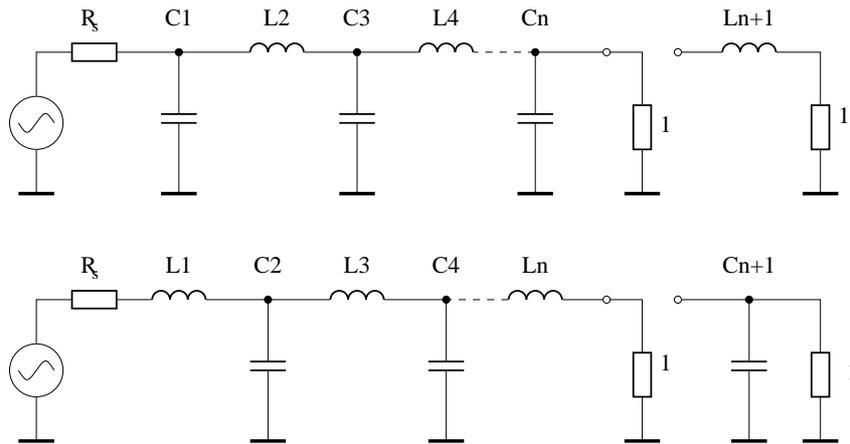}
\caption{Normalized low-pass ladder filter models}
\label{fig:LPproto}
\end{figure}

We use the tabulated values for a normalized low-pass ladder filter as
a starting point.  (Table~\ref{tab:Besselproto}, figure~\ref{fig:LPproto}).  
The filter may start with
a series inductor or a shunt capacitor, and other circuit considerations
must decide which is most appropriate.

\begin{table}[h]
 \center
 \caption{Normalized Bessel filter element values for $R_s=1$}
 \label{tab:Besselproto}
 \begin{tabular}{@{}rlllllll@{}}            
   \hline\hline
   &C1      & L2     & C3    & L4      & C5     &  L6    & C7       \\
   &L1      & C2     & L3    & C4      & L5     &  C6    & L7       \\
   \hline
  2  &0.5755  &2.1478  &       &         &        &        &          \\
  3  &0.3374  &0.9705  &2.2034 &         &        &        &          \\
  4  &0.2334  &0.6725  &1.0815 & 2.2404  &        &        &          \\
  5  &0.1743  &0.5072  &0.8040 & 1.1110  &2.2582  &        &          \\
  6  &0.1365  &0.4002  &0.6392 & 0.8538  &1.1126  &2.2645  &          \\
  7  &0.1106  &0.3259  &0.5249 & 0.7020  &0.8690  &1.1052  &2.2659    \\
  \hline\hline
 \end{tabular}
\end{table}

The element values in table~\ref{tab:Besselproto} are normalized to 
unit load resistance and unit cut-off angular frequency.  They
are to be scaled to the required impedance and frequency.  The
relations between the \emph{normalized} and the \emph{real} values for
target cut-off angular frequency $\omega$ and load impedance $Z$ are
\begin{equation}
 C_r = \frac{C_n}{\omega Z}
 \qquad \text{and} \qquad
  L_r = \frac{L_n Z}{\omega}.
 \label{eq:denormalization}
\end{equation}

\begin{figure}[h]
 \center\includegraphics[width=0.7\linewidth]{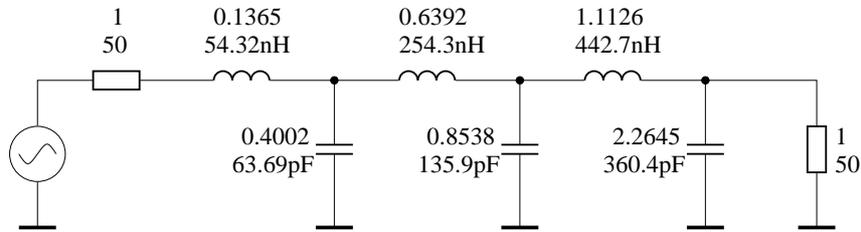}
 \caption{Schematic of a $50~\Omega$, O(6), 20~MHz low-pass Bessel filter} 
 \label{fig:BesselO6} 
\end{figure}

\begin{figure}[h]
 \center\includegraphics[width=0.6\linewidth]{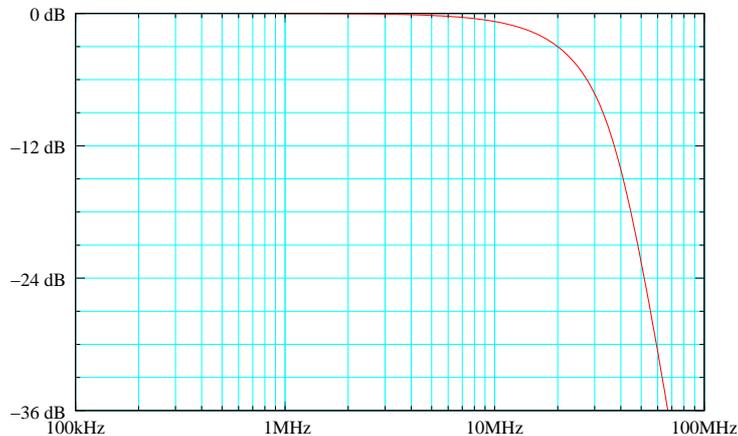}
 \caption{Frequency response of the $50~\Omega$, O(6), 20~MHz low-pass Bessel filter}
 \label{fig:BesselO6-20MHz}
\end{figure}

As an example, let's make an O(6) Bessel filter with a 20~MHz cut-off
frequency for $50~\Omega$ source and load impedances.  The angular
frequency $\omega = 2 \pi * 20~\text{MHz}$, so we find $C_r = 159.2\text{p}
\cdot C_n$ and $L_r = 397.9\text{n} \cdot L_n$. Figure~\ref{fig:BesselO6}
shows both the normalized and the scaled element values for this filter
and figure~\ref{fig:BesselO6-20MHz} shows its frequency response. Note
that the curve goes through -3~dB, its half power point, at 20~MHz,
just as intended.


Zverev lists extensive tables of element values for different
filter types and a selection of source resistances~\cite{bib:Zverev}.
These cover the great majority of filtering applications.

\subsection{Constant input impedance filters}
Constant input resistance filters do not reflect unwanted energy. This is
useful, for example, if the filter is required to terminate a long cable,
or if the source won't absorb reflected energy.  This is often the
case in beam instrumentation.  An additional feature of such filters is
that the shape of the frequency response is independent of the source
resistance.  These properties make constant input resistance filters very
useful in many applications.

The starting point for the design of a filter with constant input
resistance is the normalized lowpass prototype for zero source
impedance. This is the filter that produces the desired response with
a constant input level as a function of frequency. 

\begin{figure}[h]
 \center\includegraphics[width=0.7\linewidth]{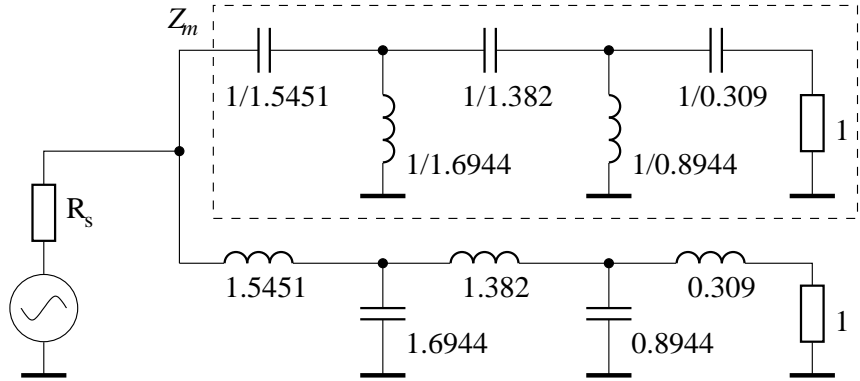}
 \caption{Constant-impedance Butterworth filter}
 \label{fig:CR-butterworth}
\end{figure}

All such filters start with a large series inductance.  Therefore,
the input impedance as a function of frequency will tend to rise above
cut-off. A shunt impedance across the filter input must be added to
maintain constant input resistance. For Butterworth filters this is
easy (Table~\ref{tab:Butterworth-rs0}). It's sufficient to add the
\emph{dual} circuit, like the example for an O(5) filter, in the dashed box
in figure~\ref{fig:CR-butterworth}.  It has inductors where the normal
filter has capacitors and vice-versa, and the component values are the
reciprocals of the values in the normal filter.
\begin{table}[h]
 \center
 \caption{Normalized element values of an O(5) Butterworth filter for $R_s=0$}
 \label{tab:Butterworth-rs0}
 \begin{tabular}{@{}rlllll@{}}            
   \hline\hline
   &L1      & C2     & L3    & C4      & L5        \\
   \hline
  5  &1.5451  &1.6944  &1.3820 &0.8944   &0.3090   \\
  \hline\hline
 \end{tabular}
\end{table}
     
\begin{table}[h]
 \center
 \caption{Normalized Bessel filter element values for $R_s=0$}
 \label{tab:Bessel-rs0}
 \begin{tabular}{@{}rlllllll@{}}            
   \hline\hline
   &L1      & C2     & L3    & C4      & L5     &  C6    & L7       \\
   \hline
  3  &1.4631  &0.8427  &0.2926 &         &        &        &          \\
  4  &1.5012  &0.9781  &0.6127 &0.2114   &        &        &          \\
  5  &1.5125  &1.0232  &0.7531 &0.4729   &0.1618  &        &          \\
  6  &1.5124  &1.0329  &0.8125 &0.6072   &0.3785  &0.1287  &          \\
  7  &1.5087  &1.0293  &0.8345 &0.6752   &0.5031  &0.3113  &0.1054    \\
  \hline\hline
 \end{tabular}
\end{table}

\begin{figure}[h]
 \center\includegraphics[width=0.8\linewidth]{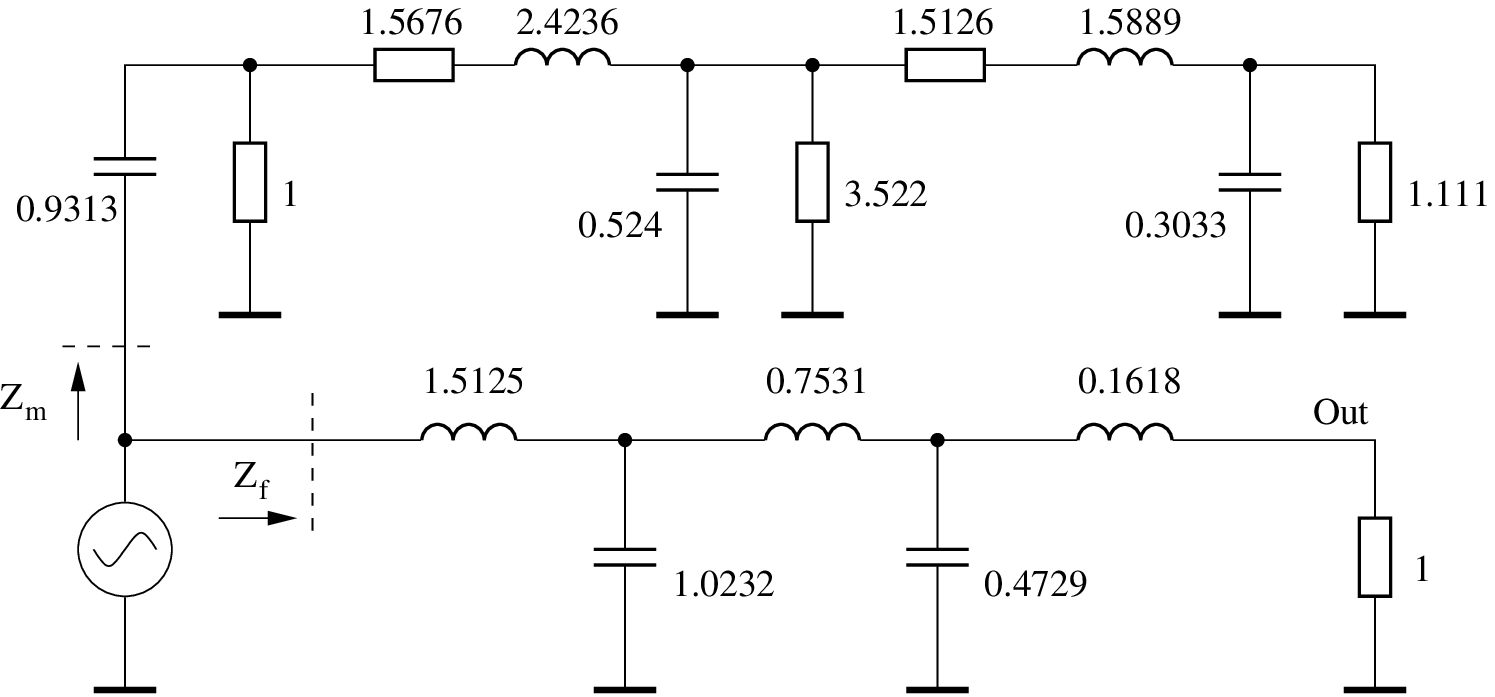}
 \caption{A constant impedance O(5) Bessel filter}
 \label{fig:CRBesselO5-2}
\end{figure}

For other filters, this simple recipe doesn't work, but it is 
sometimes still possible to derive the perfect matching impedance. 
\begin{figure}[h]
 \center\includegraphics[width=0.8\linewidth]{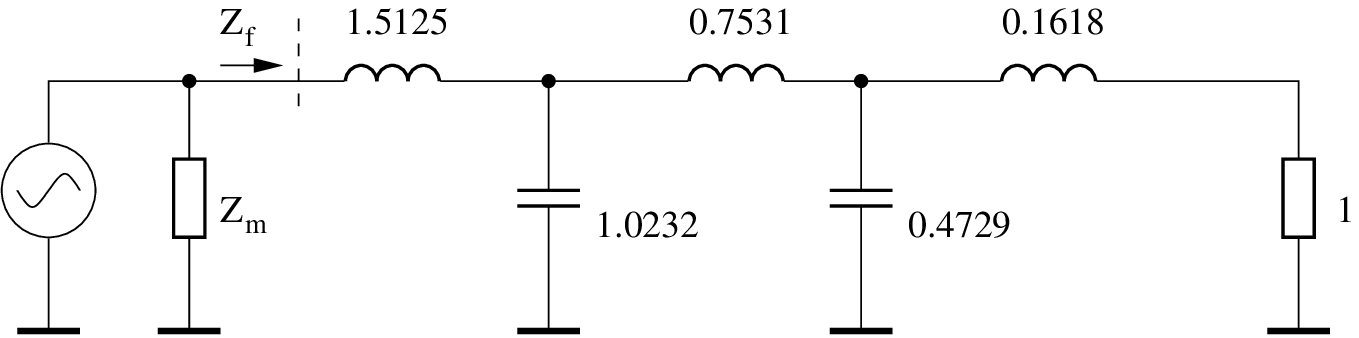}
 \caption{A constant impedance O(5) Bessel filter}
 \label{fig:CRBesselO5}
\end{figure}
For example, consider the O(5) Bessel filter of figure~\ref{fig:CRBesselO5}. This filter's
input impedance is
\begin{equation}
 Z_f = 1.5125 s + \frac{1}{1.0232 s + \frac{1}{0.7531 s + \frac{1}{0.4729 s + \frac{1}{0.1618 s +1}}}}.
 \label{eq:Zf}
\end{equation}
where $s = j\omega$.
To obtain a constant input resistance, we would have to add a 
matching shunt impedance $Z_m$ across the input such that
\begin{equation}
 \frac{1}{Z_m} + \frac{1}{Z_f}= 1 .
 \label{eq:Ym}
\end{equation}
Solving for $Z_m$, we find
\begin{equation}
 Z_m = \frac
 {1 +2.4274 s+2.61899 s^2+1.58924 s^3+0.55116 s^4+0.0891777 s^5}
 {0.9313 s+1.60635 s^2+1.22484 s^3+0.4922 s^4+0.0891777 s^5}.
 \label{eq:Ym-reduced}
\end{equation}
This can then be expanded into a continued fraction from which we can read
the component values of a ladder circuit implementing the required
shunt impedance
\begin{equation}
 Z_m = \frac{1}{0.9313s} + \frac{1}{1+\frac{1}{1.5676+2.4236s+\frac{1}{0.2839+0.524s+\frac{1}{1.5126+1.5889s+\frac{1}{0.8997+0.3033s}}}}}.
 \label{eq:Zm-expanded}
\end{equation}
A symbolic math program removes most of the tedium of working this out.
So the final (still normalized) constant-impedance filter ends up as shown
in figure~\ref{fig:CRBesselO5-2}.  Matching network values for several more
Bessel filters are listed in table \ref{tab:Besselmatch}.
Element Z2 is omitted from the table. Its value is always unity.

\begin{table}[h]
 \addtolength{\tabcolsep}{-4pt}
 \center
 \caption{Normalized Bessel filter matching circuit element values}
 \label{tab:Besselmatch}
 \begin{tabular}{@{}rllllllll@{}}            
   \hline\hline
   &Z1          & Z3              & Z4             & Z5            &  Z6           & Z7           &Z8             \\
   \hline
  3  &0.913F   &1.755+2.428H   &2.428F          &               &               &              &               \\
  4  &0.925F   &1.570+2.407H   &2.413//0.5469F  &3.240+1.447H   &               &              &               \\
  5  &0.931F   &1.568+2.424H   &3.522//0.5240F  &1.513+1.589H   &1.111//0.303F  &              &               \\
  6  &0.935F   &1.616+2.446H   &4.276//0.5012F  &1.009+1.679H   &2.514//0.330F  &3.558+0.945H  &               \\
  7  &0.936F   &1.669+2.465H   &4.647//0.4838F  &0.777+1.709H   &4.004//0.347F  &1.656+1.110H  &1.131//0.191F  \\
  \hline\hline
 \end{tabular}
\end{table}

\begin{figure}[h]
 \center\includegraphics[width=.8\linewidth]{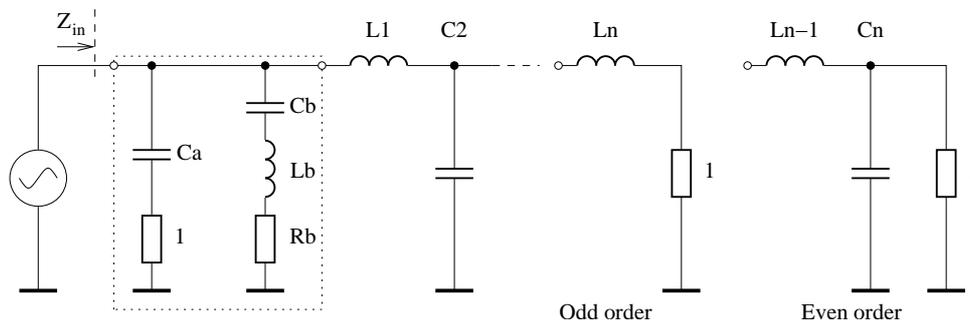}
 \caption{Normalized constant resistance low-pass ladder filter prototype}
 \label{fig:crfilter}
\end{figure}

Although this works for Bessel filters of any order, for orders greater than three,
there is a simpler solution which works just as well in practice.
A series RC circuit across the filter input can restore the input
impedance to unity for very high frequencies, and a series RLC resonator
can be positioned over the transition region to minimize the impedance
bump (Fig.~\ref{fig:crfilter}, box).

\begin{table}[h]
 \addtolength{\tabcolsep}{-1pt}
 \center
 \caption{Element values for some normalized Bessel constant input resistance filters}
 \label{tab:crBessel}
 \begin{tabular}{@{}llllllllllll@{}}
  \hline\hline
    &Ca      &Cb     &Lb     &Rb     &L1     &C2     &L3     &C4     &L5     &C6     &L7\\
  \hline
  3 & 0.5804 &0.3412 &0.9915 &2.6161 &1.4631 &0.8427 &0.2926 &       &       &       &\\
  4 & 0.6121 &0.3143 &1.0646 &2.7036 &1.5012 &0.9781 &0.6127 &0.2114 &       &       &\\
  5 & 0.6465 &0.2834 &1.1613 &2.8896 &1.5125 &1.0232 &0.7531 &0.4729 &0.1618 &       &\\
  6 & 0.6622 &0.2683 &1.2094 &3.0029 &1.5124 &1.0329 &0.8125 &0.6072 &0.3785 &0.1287 &\\
  7 & 0.6876 &0.2452 &1.2955 &3.2070 &1.5087 &1.0293 &0.8345 &0.6752 &0.5031 &0.3113 &0.1054\\
  \hline\hline
 \end{tabular}
\end{table} 

\begin{figure}[h]
 \center\includegraphics[width=0.8\linewidth]{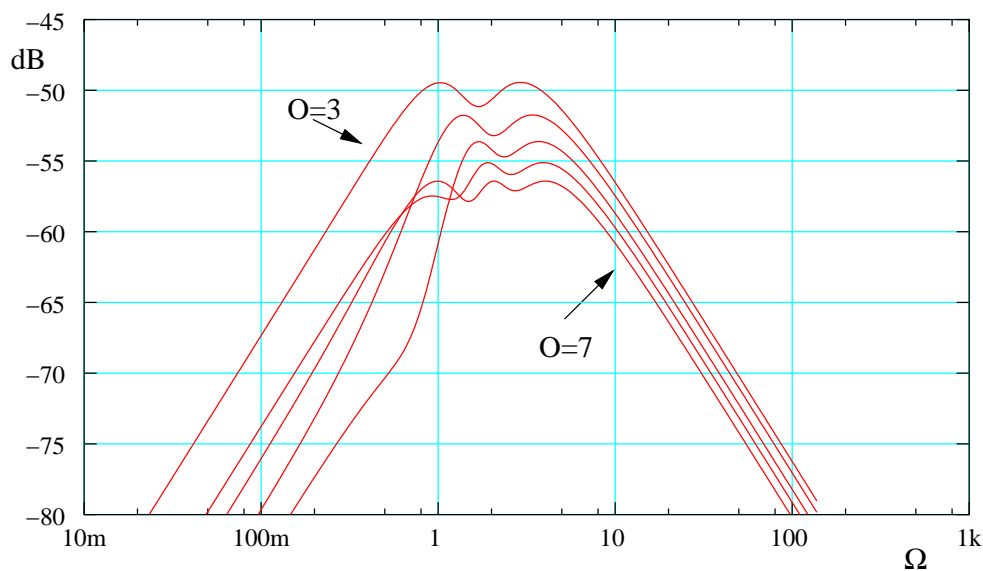}
 \caption{Reflection coefficients of constant resistance Bessel filters vs. normalized frequency}
 \label{fig:BesselS11}
\end{figure}

As can be seen in figure~\ref{fig:BesselS11}, the calculated reflection
coefficient is nearly everywhere below $-50$~dB, meaning that
$Z_{in}$ never strays by more than about 0.3\% from unity.  Although
this is not perfect, it is still
much better than what can actually be achieved with realistic component
tolerances.
Table~\ref{tab:crBessel} lists the element values for some Bessel filters
with the associated matching circuits. Moreover, this can be done for some
other filter types too.
Element values for Bessel filters up to O(10), as well as for Gaussian
and some equi-ripple phase error filters are given in~\cite{bib:crfilter}.

To summarize, the exact matching network obtained by solving Eq.~(\ref{eq:Ym})
is fine for low order Bessel filters, while the simplified matching
network of figure~\ref{fig:crfilter} is preferred for higher orders, as
well as for Gaussian or equi-ripple filters. Either way, the quality of
the match is limited by the tolerances of available components.

\subsection{Band-pass filters}
\begin{figure}[h]
 \center\includegraphics[width=0.8\linewidth]{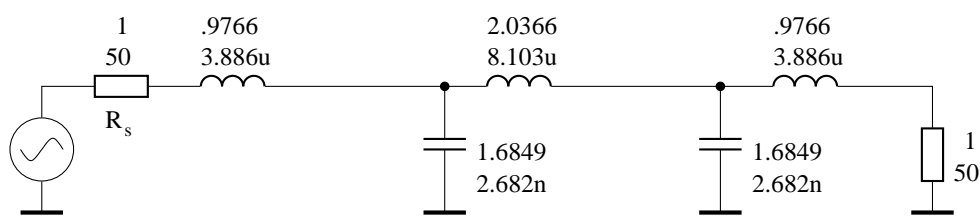}
 \caption{Normalized~(upper values) and scaled~(lower) Chebyshev O(5) low-pass}
 \label{fig:chebyshev5}
\end{figure}

The low-pass prototype filter tables can also be used to design band-pass
filters. You start off by calculating a low-pass filter with the target
\emph{bandwidth} as cut-off frequency. You then replace each series branch
by a series L-C combination and each shunt branch by a parallel L-C,
both tuned to the desired centre frequency. As and example, let's
design an Chebyshev O(5) band-pass with a 2~MHz bandwidth and a 
20~MHz centre frequency.
Table~\ref{tab:chebyshev5}
and figure~\ref{fig:chebyshev5} show the prototype low-pass element values.
\begin{table}[h]
 \center
 \caption{Chebyshev O(5) low-pass prototype values}
 \label{tab:chebyshev5}
 \begin{tabular}{@{}lllll@{}}            
  \hline\hline
  L1      & C2     & L3    & C4      & L5      \\
  \hline
  0.9766  &1.6849  &2.0366 & 1.6849  &0.9766 \\ 
  \hline\hline
 \end{tabular}
\end{table}

\begin{figure}[h]
 \center\includegraphics[width=0.8\linewidth]{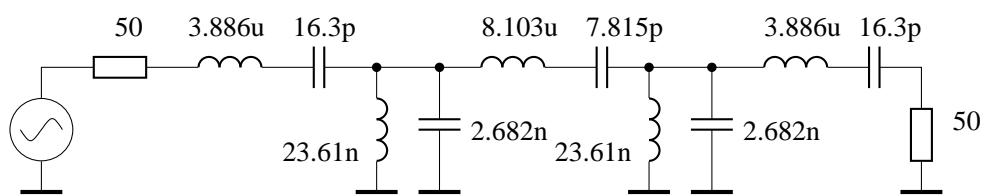}
 \caption{Chebyshev O(5) band-pass centered on 20~MHz}
 \label{fig:ChebyO5-BP}
\end{figure}

\noindent
First we scale the filter to a low-pass with a 2~MHz cut-off frequency (Fig~\ref{fig:chebyshev5}).
Then we resonate all elements to 20~MHz, that is, we pair up each L with a C
and vice-versa, such that $ 1/(2 \pi \times 20~MHz) = \sqrt{L C}$ (Fig.~\ref{fig:ChebyO5-BP}).
\begin{figure}[h]
 \center\includegraphics[width=0.8\linewidth]{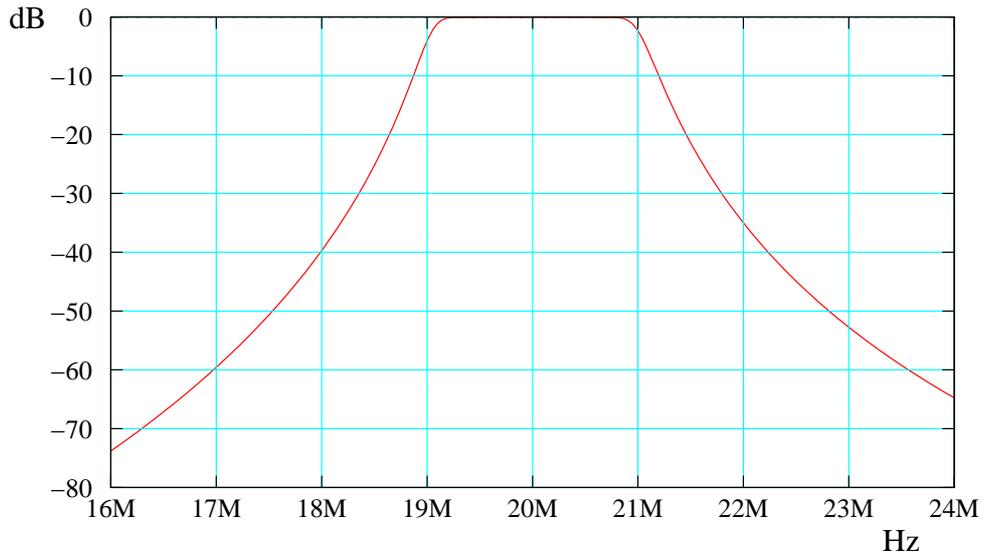}
 \caption{Chebyshev O(5) band-pass frequency response}
 \label{fig:ChebyO5-BP-response}
\end{figure}
The frequency response of the resultant bandpass filter is shown in
figure~\ref{fig:ChebyO5-BP-response}.
Be warned that this recipe can easily lead to awkward component values.
For example, in the filter of figure~\ref{fig:ChebyO5-BP}, it may be difficult to
find or make an inductor of 8.1~$\mu$H that is still inductive around
20~MHz. Several circuit transforms exist that
may help the situation~\cite{bib:Rhea2001}. For very wide bandpass 
filters, it may be easier to cascade a low-pass and a high-pass. For
very narrow filters, coupled resonator filters are more appropriate.
Methods to design those are outlined in \cite{bib:Bowick1982}.


\subsection{Constant resistance T and L networks}
\label{ssec:crn}
These little networks can be inserted into matched systems, because
provided they are terminated into their characteristic impedance $R$,
they present a flat input impedance, independent of frequency.  They are
handy as frequency response tweaks for cables and amplifiers, input
impedance correcting elements for out-of-band signals and more. They
can be cascaded without interaction. However, you can't use them
reproduce the responses of the L-C filters in the previous sections,
because they can't have complex conjugate pole pairs
~\cite{bib:Balabian1969}.

$Z_a$ and $Z_b$ are complex impedances such that $Z_a Z_b = R^2$.
The frequency response of the networks is then
\begin{equation}
 H(f) = \frac{R}{R+Z_a}. 
 \label{eq:crn}
\end{equation}
\begin{figure}[h]
 \begin{minipage}[b]{0.3\linewidth}
  \includegraphics[width=0.8\linewidth]{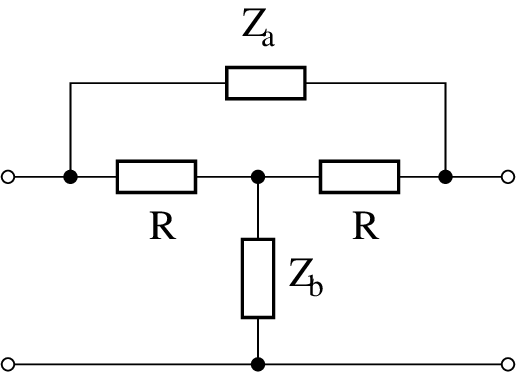}
  \caption{Bridged T}
  \label{fig:bridged-T}
 \end{minipage}
 \hfill
 \begin{minipage}[b]{0.3\linewidth}
  \includegraphics[width=0.8\linewidth]{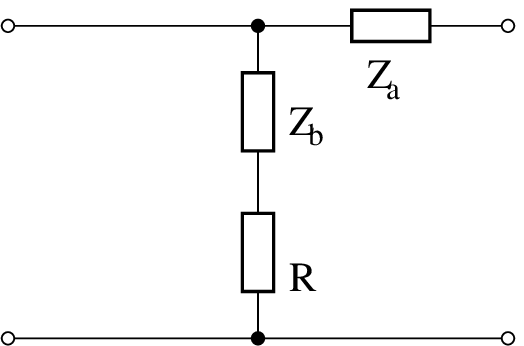}
  \caption{Right L}
  \label{fig:right-L}
 \end{minipage}
 \hfill
 \begin{minipage}[b]{0.3\linewidth}
  \includegraphics[width=0.8\linewidth]{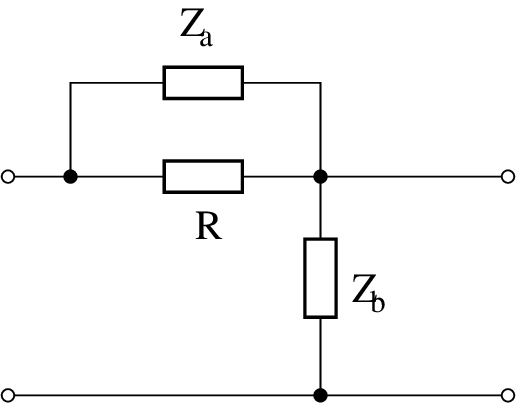}
  \caption{Left L}
  \label{fig:left-L}
 \end{minipage}
\end{figure}

\begin{figure}[h]
 \begin{minipage}[b]{0.5\linewidth}
  \center\includegraphics[width=0.5\linewidth]{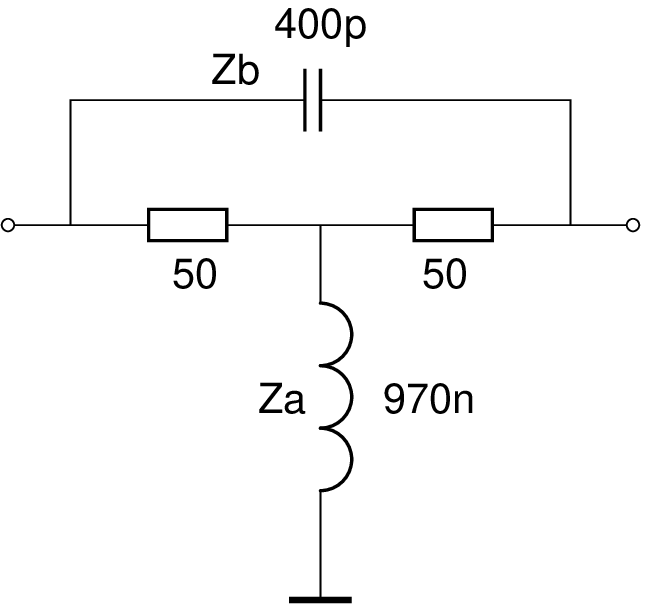}
  \caption{Bridged-T constant resistance test jig}
  \label{fig:jig}
 \end{minipage}
 \begin{minipage}[b]{0.5\linewidth}
  \center\includegraphics[width=0.9\linewidth]{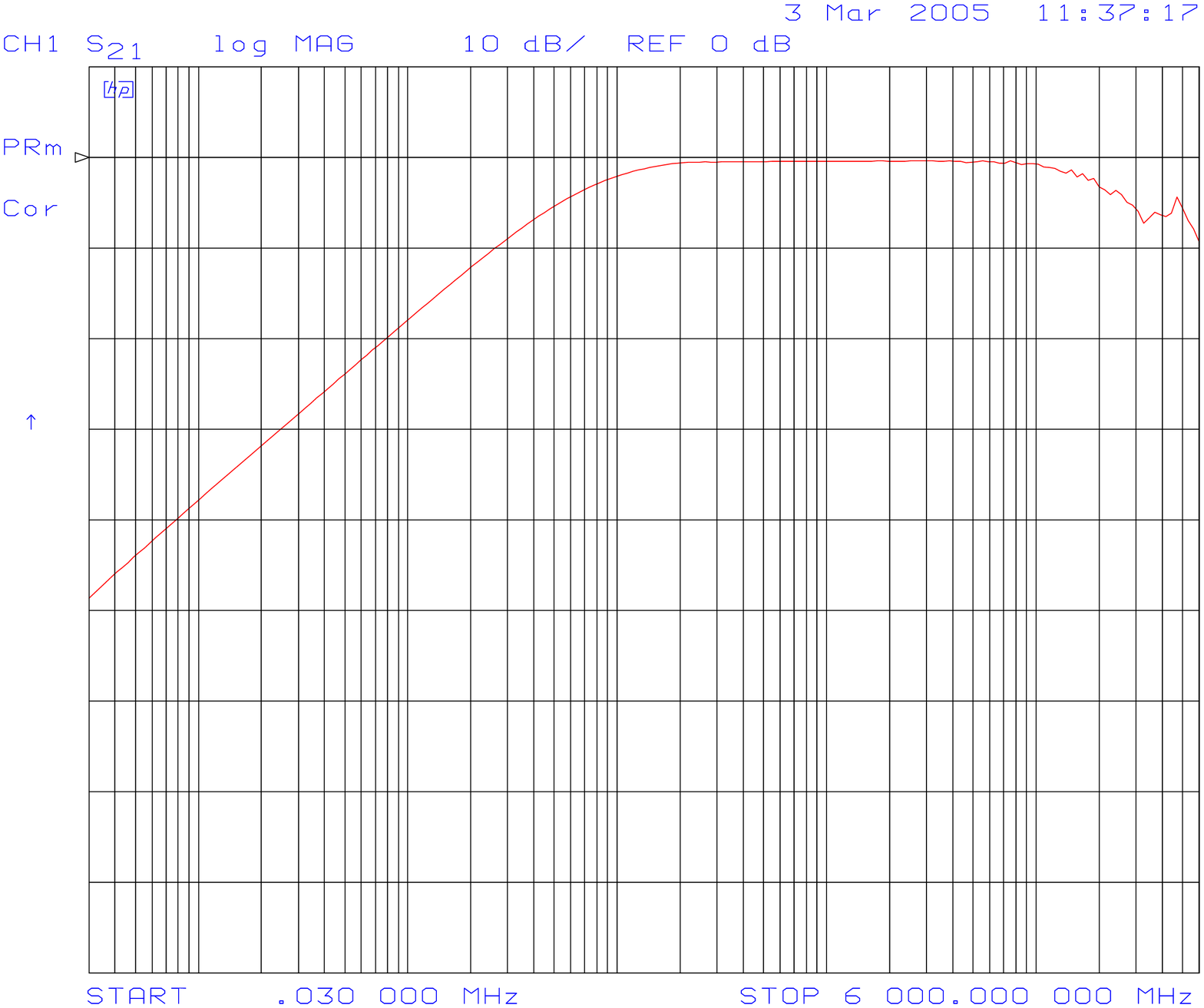}
  \caption{Test jig frequency response}
  \label{fig:jigresponse}
 \end{minipage}
\end{figure}

An example application, figure~\ref{fig:jig}, is a test jig simulating
the high-pass characteristic of a capacitive beam position pick-up in
a pre-amplifier test setup.

\noindent
Examples of other applications are described in \S\ref{sssec:lna} and in \cite{bib:Amsel1971}.          

\section{Electronic noise}

\begin{wrapfigure}{R}{0.3\textwidth}
\center\includegraphics[width=.7\linewidth]{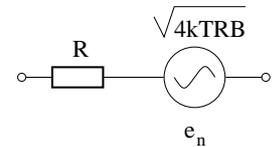}
\caption{Resistor noise model}
\label{fig:Rnoise}
\end{wrapfigure}
Noise, the undesirable random variations of voltage or current, is the
ultimate limit that prevents us from appreciating signals in infinite
detail. We consider \emph{noise} the random fluctuations of currents and
voltages inherent in the circuit components. In this context, noise coming
from unspecified outside sources is \emph{interference}.  There are two
fundamental sources of noise: Thermal noise and shot noise.

Thermal or Johnson noise is the result of the continuous thermal agitation
of the charges in a conductor~\cite{bib:Johnson1928}.  Any device
that converts electrical power into heat --think 'resistor'-- also does
the opposite. A purely reactive impedance does not generate noise. The
electrical noise power density in W/Hz available at the terminals of a
simple resistor is:
\begin{equation}
P = k T,
\end{equation}
with $k$ Boltzmann's constant 13.8~yJ/K and $T$ the absolute temperature.
The probability distribution of the noise voltage is Gaussian.
This noise is 'white', by which we mean that the power spectral density
is constant over frequency.
(At least in the frequency and temperature ranges covered by ordinary
electronics. At very high frequency or at very low temperature, things
change.)

An ordinary resistor can thus be modelled as having a series noise
voltage source of value
\begin{equation}
e_n = \sqrt{4k T R B},
\label{eq:johnsonnoise}
\end{equation}
with R the resistance in Ohms and B the bandwidth in Hertz.  As a point of
reference, the common 50~$\Omega$ termination resistor at room temperature
has a built-in noise source of about 1~nV/$\sqrt{\text{Hz}}$.  It is a
small value, but quite often it isn't small enough! In addition, some
resistor types --carbon composite resistors for example-- are notorious
for producing more noise than that.

Shot noise, also Schottky noise is what happens when an electric current
is made to flow across a potential barrier, such as a semiconductor
junction or a vacuum gap~\cite{bib:schottky1926}.  It is the
consequence of the fact that current is composed of discrete charges.
The current then flows in discrete lumps of one fundamental charge
$q_0 = 160~\text{zC}$. As a result, the current varies with a standard
deviation of
\begin{equation}
i_n = \sqrt{2 q_0 I B}.
\end{equation}
This noise is also white and Gaussian (Except for very small currents).
In metallic conductors, like wires or even resistors, where long range
correlations between charge carriers exist, shot noise is very much less.

\subsection{Noise factor, noise figure}
The noise factor $F$ is a figure
of merit often specified for amplifiers and individual components like
transistors~\cite{bib:haus1963}.         
Basically, it's the ratio of the total noise of an amplifier to the
portion contributed by the source resistance alone. 
Obviously, its value can never be less than unity.
\begin{equation}
 F = \frac{4 k T R_s + {v_n}^2}{4 k T R_s}
 \label{eq:NF}
\end{equation}
\begin{wrapfigure}{R}{0.4\textwidth}
 \center\includegraphics[width=0.8\linewidth]{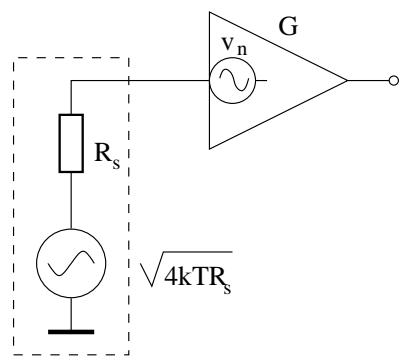}
 \caption{Noise factor model circuit}
 \label{fig:NF}
\end{wrapfigure}
The value is often expressed in decibels and is then called
the noise figure: $\mathit{NF} = 10 \log{F}$.

The source impedance is often not specified. Depending on
context, it may be the standard 50~$\Omega$ of common RF
measurement instruments, or it may be some optimum value
that makes the amplifier look good. 

\subsection{Measuring noise in amplifiers}
Every resistor, every semiconductor contributes some noise to pollute
the signal being amplified.  Noise is usually measured at the amplifier's
output. However, by convention, the noise level of an amplifier is always
specified referred to its input.  In practice, that is indeed also where
most of the noise usually comes from.

The natural way to find the input-referred noise contribution of an
amplifier would be to divide the measured output noise by the gain,
giving the quantity in the numerator of Eq.~(\ref{eq:NF}), and to then
subtract the contribution of the source resistance. This is fraught
with pitfalls.  First of all, it is hard to measure absolute noise
power levels.  RF power meters have various ways of detecting signal
levels and are often calibrated to display correct power levels only for
single-frequency sine wave signals. It's not always clear how they behave
with broadband Gaussian noise. The effective measurement bandwidth is also
often uncertain and the response may not be flat over that bandwidth.
The amplifier's gain also needs to be accurately known, and it may not
be flat over the measurement device's bandwidth either.  Finally you
have to subtract the source noise.

All this severely affects the accuracy of the result.  In fact, it's not
at all unusual to end up with negative values for the amplifier's own
noise contribution, which would be, of course, nonsense. Fortunately,
there are better ways.


\subsection{The Y-method}
The Y-method consists in connecting two different noise sources with
known levels to an amplifier and measuring the resultant change
of output noise~\cite{bib:haus1963}.  The noise sources may be simple
resistors kept at different temperatures, for which the noise levels
are known from first principles, Eq.~(\ref{eq:johnsonnoise}), or they may be
calibrated noise sources sold for that purpose by reputable instrument
manufacturers.
\begin{figure}[h]
\center\includegraphics[width=0.5\linewidth]{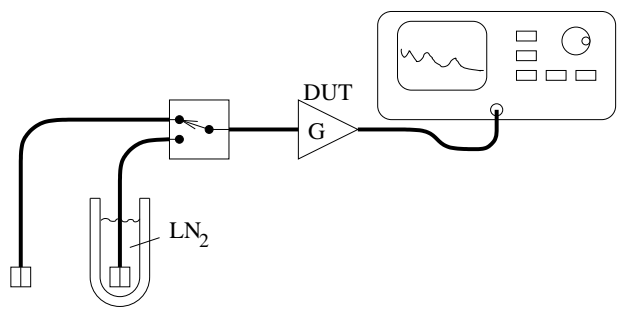}
\caption{Measuring noise using the Y-method}
\label{fig:Ymethod}
\end{figure}
The lower the noise
contribution of the amplifier under test, the closer the ratio of output
power levels will approach the ratio of the input noise levels.

The following argument assumes impedance-matched amplifiers.
Let's suppose we use two equal-valued resistors with $R=Z_0$ at different
temperatures.
Let $P_a$ be the input noise power density of the amplifier,
$P_h$ the power density of the hot source and $P_c$ the power
density of the cold source. All noise sources are uncorrelated.
The two equations below then give the 
power density at the output of the amplifier with the hot,
respectively cold source connected:
\begin{equation}
\begin{aligned}
P_1 = G(P_a + P_h),\\
P_2 = G(P_a + P_c).
\end{aligned}
\label{eq:Pn}
\end{equation}

While it's not easy to accurately measure absolute noise power levels,
measuring the ratio of two such powers is rather simple. It requires
neither an accurate absolute calibration of the measuring instrument,
nor exact knowledge of the DUT gain and frequency response. A spectrum
analyzer can easily detect the change in noise level in the amplifier's
output due to a switch from the hot to the cold source, provided the DUT
gain is high enough to make the spectrum analyzer's own noise contribution
negligible. Note that spectrum analyzers often have a 20~dB
attenuator at the input, so that 
30~dB or more of DUT gain may be needed to fulfil that condition.

So let's define the Y-factor as the ratio of the power values defined in
Eq.~(\ref{eq:Pn}):
\begin{equation}
Y = \frac{P_1 }{P_2 } = \frac{G(P_a + P_h) }{G(P_a + P_c) }.
\end{equation}
The gain $G$ of the amplifier drops out right away. Solving for $P_a$ yields
\begin{equation}
P_a = \frac{P_h - Y P_c }{Y-1}.
\end{equation}
Finally, we can express the amplifier's noise as an effective
temperature $T_n = P_a/k$ or as a
voltage noise density in $\text{V}/\sqrt{\text{Hz}}$:
\begin{equation}
V_n = \sqrt{P_a R} = \sqrt{k T_a R}.
\end{equation}

\begin{figure}[h]
\center\includegraphics[width=0.6\linewidth]{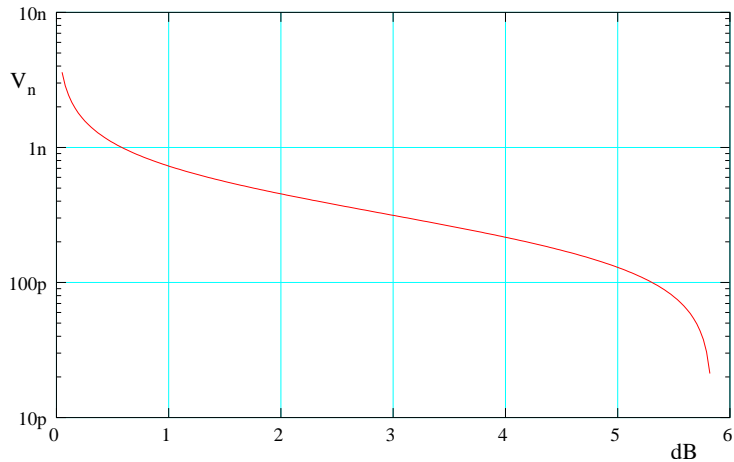}
\caption{Input referred noise voltage density vs. Y-factor in dB}
\label{fig:Ymethodplot}
\end{figure}
For very quiet amplifiers, we might use one resistor at 
room temperature and another immersed in liquid $N_2$ , at 77~K
(Fig.~\ref{fig:Ymethod}).
For a hypothetical noiseless amplifier, the Y-factor would be
just the ratio of noise source temperatures,
293/77 = 3.8, corresponding to 5.8~dB, easily seen on a
spectrum analyzer display with some averaging. 
Note that to get reasonably accurate results, the noise levels of
the sources should be of a similar magnitude as that of the
amplifier under test. This is apparent in figure~\ref{fig:Ymethodplot},
which plots the equivalent input noise voltage density against the
Y-factor for the setup of figure~\ref{fig:Ymethod}
and where the slope of the curve increases at both ends.
For higher noise levels, several instrument manufacturers propose  
calibrated noise sources with apparent noise temperatures in the
10~kK range.

\subsubsection{Noise in matched amplifiers}

\begin{figure}[h]
 \begin{center}
  \includegraphics[width=0.6\linewidth]{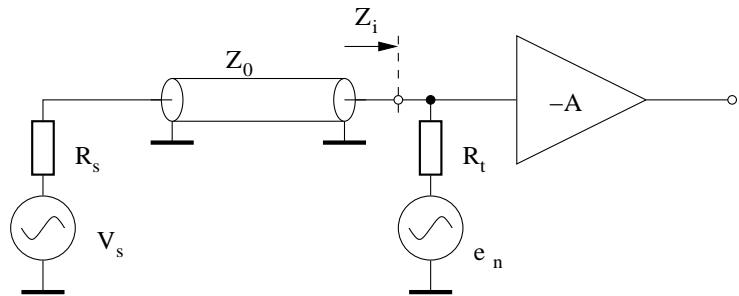}
  \caption{Termination with a simple resistor $R_t=Z_i$ to ground}
  \label{fig:term1} 
 \end{center}
\end{figure}

If an amplifier is connected to a passive filter or a coaxial cable,
it must have the correct input impedance, usually $50~\Omega$.
If this is done with a simple resistor $R_t = 50~\Omega$ to ground
(Fig.~\ref{fig:term1}), the thermal noise of that resistor may turn out
to be the main contributor to the total noise.

\begin{equation}
V_n = \sqrt{4kTR_t}\left(\frac{R_s}{R_s+Z_i}\right) = \sqrt{kTZ_i} 
\label{eq:Vn1}
\end{equation}

\begin{figure}[h]
 \begin{center}
  \includegraphics[width=0.6\linewidth]{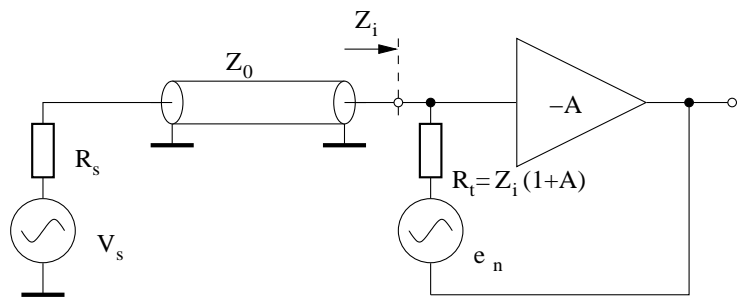}
  \caption{Termination using feedback from the output}
  \label{fig:term2}
 \end{center}
\end{figure}

It's possible to do better if we design the amplifier to have
a largish negative gain $-A$, and to then use a feedback resistor from
the output back to the input to set the terminating impedance 
(Fig.~\ref{fig:term2}).

To keep the value of input impedance $Z_i$ the same, $R_t$ must now
be much larger: $R_t = (1+A) Z_i$.  The input referred noise voltage
density due to the termination is now
\begin{equation}
 V_n =  \sqrt{\frac{kTZ_i}{1+A}}, 
 \label{eq:Vn2}
\end{equation}
much less than in Eq.~(\ref{eq:Vn1}).  This is how a matched amplifier can
have a very low apparent noise temperature, even though the whole circuit
is kept at ordinary room temperature.
However, phase shifts and gain errors in the amplifier will affect $Z_i$,
possibly to the point of making the input impedance go negative
at some frequencies, causing instability with reactive sources.
Additional measures must be taken to prevent that, for example
by inserting constant-impedance low-pass elements (See~\S\ref{ssec:crn}).

\subsubsection{An example of a low-noise amplifier stage}
\label{sssec:lna}
\begin{figure}[h]
\center\includegraphics[width=0.8\linewidth]{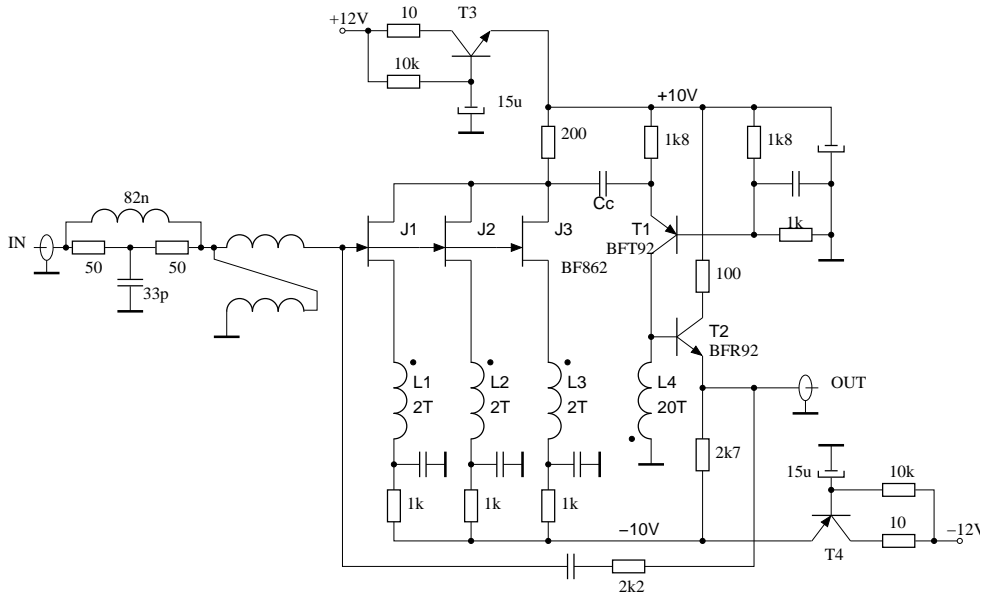}
\caption{A low-noise amplifier example}
\label{fig:LNA}
\end{figure}

The design uses a JFET (BF862) input stage (Fig.~\ref{fig:LNA}).
The voltage noise of the BF862 is specified as
$V_n=0.8$~nV/$\sqrt{\text{Hz}}$.  These being JFETs, the noise voltage
is largely dominant over the contribution of the current noise at low source impedances,
so we trade off one against the other by putting three JFETs in parallel,
which brings $V_n$ down to 460~pV/$\sqrt{\text{Hz}}$.  This also triples
the parasitic capacitances, of which especially $C_{GD}$ is troublesome,
because its apparent value is multiplied by the gain of the JFETs (Miller
effect). The deleterious effects of $C_{GD}$ are limited by cascoding
the JFETs with a BFT92 PNP transistor (T1). 

The three 1~k$\Omega$ resistors in the source leads of the JFETs
set the DC current in each to 10~mA, which is the minimum
guaranteed $I_{DSS}$ for this type. The transfer admittance $y_{fs}$
at this current is about 35~mS per JFET, so 105~mS altogether. 
The collector load of the BFT92 is essentially the magnetization
inductance of L4, about 5~mH. 
The open loop gain is then $A=0.105*j\omega (L4)$.

A final emitter follower NPN transistor T2 provides a sufficiently low
output impedance to drive further stages.  Gain-setting feedback is
through the transformer composed of L1 to L4. This avoids the thermal noise
that would have been introduced by feedback resistors. The turns ratio is 10,
and so the gain in the feedback path is $\beta = 0.1$.  The closed-loop
gain is then $G = \frac{-A}{1+\beta A}$, where $A$ is the open-loop gain.
With $A$ appreciably larger than 10 above 10 kiloherz, the overall
amplifier gain $G \approx -1/\beta = -10$. Apart from yielding a flat,
well defined gain over  a large range of frequency, this feedback also
considerably improves the amplifier's linearity.

The transformer core is a tiny high-permeability amorphous metal toroid
core that would saturate with only a few mA-turns of current.  The DC
bias current in the cascoding PNP transistor has been chosen to cancel
the DC magnetization of the core due to the DC current of the JFETs.

The input impedance is set to $200~\Omega$ by feedback through the 2k2
resistor. 
A 1:4 transmission line transformer brings that down to $50~\Omega$,
doubling the gain and halving the input-referred voltage noise at the same time.
A constant resistance bridged-T low-pass section hides the 
excursions of the input impedance near the high-frequency cut-off,
keeping the amplifier stable at the same time.

The final result is an amplifier with 26~dB of gain, a bandwidth of
10~kHz to 75~MHz and a noise level of 260~pV/$\sqrt{\text{Hz}}$.
This corresponds to an effective noise temperature of the amplifier
of about 30~K, despite the fact that the amplifier is at room
temperature.

The power supply lines are filtered using capacitance multipliers (T3, T4).
Ordinary linear voltage regulators have noise levels
of the order of 0.003\% of the output voltage in a 10~kHz bandwidth,
which seems very good until you realize that this is several thousand
times the target noise level of the amplifier, and that this circuit
does not by itself have a very good rejection of power supply noise.
Capacitance multipliers reduce the noise density of the power supplies
to more acceptable levels, of the order of a few nV/$\sqrt{\text{Hz}}$.

\subsubsection{On the edge between noise and interference}
Johnson and Schottky noise are fundamental noise sources. There are
several more noise-like phenomena lurking in the dark.  It depends on
the application requirements if any of these are important. They often
aren't.  Here are some things to keep in mind, in no particular order
and without going into details.  Temperature fluctuations caused by fans
or turbulent airflow affect component values through their temperature
coefficients. The Seebeck or thermoelectric effect causes small voltages
to be developed when different metals are joined and are subjected to
thermal gradients. Resistors and capacitors change value under mechanical
stress and provide a path for acoustic noise to find its way in. Moreover,
high-value ceramic capacitors may be piezo-electric. Carbon and cermet
resistors suffer from excess noise, that is, noise over and above the
Johnson noise, which manifests itself when a voltage is applied.

Semiconductors have 1/f noise, which becomes increasingly important
at low frequency.  Low noise transistors and integrated circuits
usually have a 1/f corner frequency specification, below which this
noise dominates all other sources. Semiconductors are also sensitive
to light. Beware of components in translucent packages, such as glass
diodes or ceramic-packaged transistors or ICs. Zener diodes above about
5 volts are actually avalanche diodes, which produce lots of noise,
neither white nor Gaussian.

\subsection{Grounding and Shielding, Interference}

Interference can really spoil your day. Often it involves circuit
elements that are assumed to be negligible such as the resistance of
ground connections or cable screens, or the parasitic inductance of wires,
resistors and capacitors.

It may be difficult to get a good model of the way interference couples
into the signal. Usually, when the relevant coupling mechanism and the
associated circuit elements have been identified, the solution to many
EMC problems will be relatively straight-forward~\cite{bib:ott1988, bib:morrison1998}.

Three coupling mechanisms need to be considered for
interference mitigation. In no particular order:

  - Common impedance coupling.
  High current paths should not share a conductor with low-level signals.
  Bear in mind that even a thick short straight wire has resistance
  and inductance. A sizable voltage can appear across its ends if
  the current is big enough or varies fast enough.
  \begin{figure}[h]
   \begin{minipage}[b]{0.45\linewidth}
    \center\includegraphics[width=0.9\linewidth]{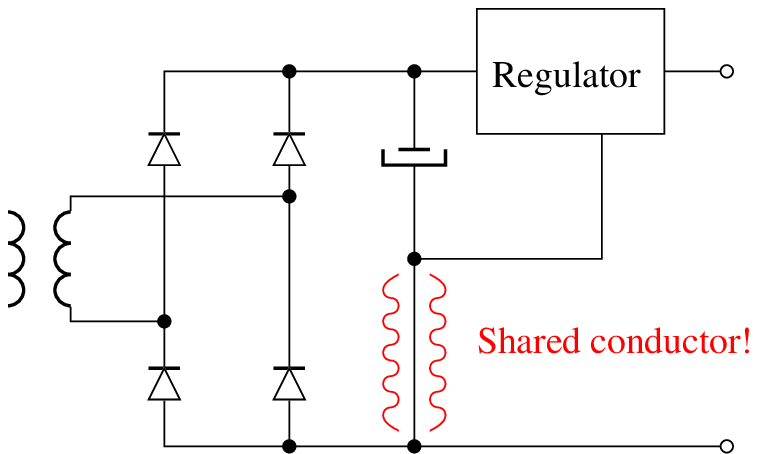}
    \caption{Common impedance coupling!}
    \label{fig:supply-a}
   \end{minipage}
   \begin{minipage}[b]{0.55\linewidth}
    \center\includegraphics[width=0.65\linewidth]{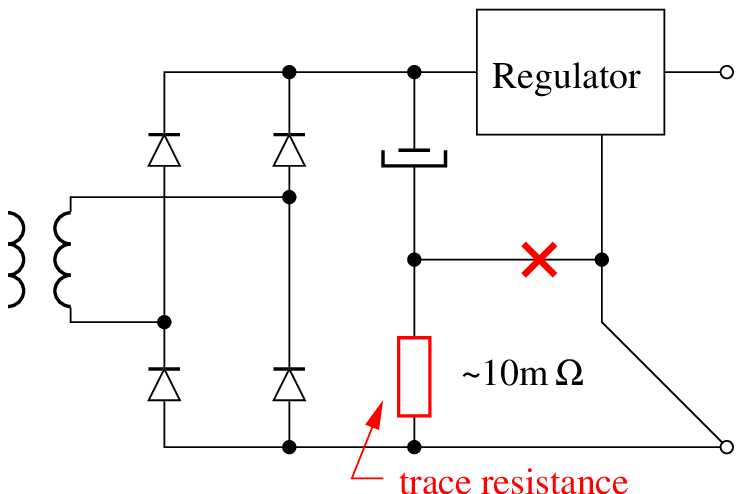}
    \caption{Solution to remove common impedance coupling}
    \label{fig:supply-b}
   \end{minipage}
  \end{figure}
  Consider the simplified power supply schematic of
  figure~\ref{fig:supply-a}. The ground reference
  of the regulator shares a conductor with the rectifier reservoir
  capacitor. This wire or trace carries large current impulses which
  cause the regulator reference to jump, and the output voltage with it.
  Connecting the reference to the supply's negative output terminal 
  instead will result in a much quieter output voltage
  (Fig~~\ref{fig:supply-b}).

  Star-point wiring, where all connections to a node meet at one point,
  is a common solution to  common impedance coupling problems.  However,
  it works well ony for low frequency.  A full conductive plane works much
  better. Possibly, some slots or a peninsula may keep large currents away
  from sensitive areas, but you should avoid cutting up ground or power
  planes unnecessarily. Avoid running wires or PCB traces across slots.
  
  - Inductive coupling. Current flows in closed loops. Minimize the area
  of loops with high $dI/dt$. Keep wires with direct and return current
  close together. Use local bypass capacitors. Also identify nearby
  loops involving low-level signals. Keep those small too.  Increase the
  distance between victim and aggressor loops.  Magnetic shielding may help if
  the shield is correctly chosen and oriented.  Static and low-frequency
  magnetic fields can be shielded with soft magnetic materials parallel
  to the field direction. High frequency fields are better shielded with
  eddy current shields normal to the magnetic field.
 
\begin{figure}[h]
   \center\includegraphics[width=0.4\linewidth]{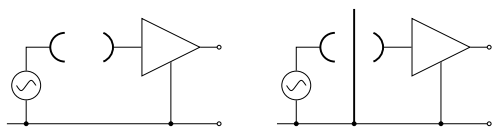}
   \caption{Capacitive coupling and electrostatic screening}
   \label{fig:Ecoupling}
  \end{figure}

  - Capacitive coupling (Fig.~\ref{fig:Ecoupling},~left). Look for nodes
  with high $dV/dt$. Keep those nodes small and close to a ground plane,
  to confine the fields. Put grounded screens around them. Reduce
  $dV/dt$ if possible.  Also look for high impedance nodes, such as
  amplifier inputs and the like. Keep those nodes compact and far from
  the previous kind. Reduce the impedance, if possible.  Keep them close
  to ground. Shield them (Fig.~\ref{fig:Ecoupling},~right).

\subsection{Radiation damage}
Radiation affects mostly
semiconductors~\cite{bib:Dentan2000}. The passive components are
usually plenty radiation hard, except, maybe surprisingly, cables
and connectors. Most contain PTFE, which is    
is a low-loss dielectric and doesn't melt at soldering temperatures.
Unfortunately, halogenated plastics deteriorate rapidly under irradiation
and then exude corrosive emanations that attack nearby metal surfaces.
Poor electrical contacts result. Paradoxically, it may be preferable
to use connectors of the 'cheap and nasty' variety, using polyethylene or
polystyrene dielectrics, which hold up much better under irradiation.

The basic mechanism by which ionizing radiation damages semiconductors
is that it dislodges atoms in the crystal structure, creating increased
opportunities to scatter charge carriers, leading to an increased
probability of recombination. The result is that bipolar transistor current gain
is reduced progressively, mostly at low bias current levels.

There is no real substitute for measurement, but some sweeping
observations can be made.
Some transistors are more affected than others. Gold-doped transistors
withstand many kilograys without deteriorating appreciably.
General purpose small-signal transistors lose current gain exponentially,
with decay constants in the 1~kGy ballpark.
Lateral PNP transistors, used inside ICs, fail at low doses, because they already have
low current gain to begin with and in addition, a large base region in
which to capture damage. Such transistors are still used in a
number of popular operational amplifiers and voltage regulators.

JFETs suffer increased leakage current of the gate junction. In MOSFETs,
radiation ejects electrons from atoms in gate insulation layers, leaving
trapped positive ions behind and causing a downward shift in threshold
voltage.
All these effects tend to reduce amplifier gain, increase noise and
upset bias conditions.

The obvious path to rad-hard equipment design is to install as little as
possible of the electronics in the irradiated area and to place what's
left out of harm's way if at all possible.  Other strategies are, of
course, a judicious choice of components, to design your circuit to have
largish standing currents and to tolerate large ranges of current gain
or threshold voltage. An ample gain reserve, combined with the liberal
application of feedback, helps to stabilize bias points and gain values.

In this context, note that older ICs are often more resilient because
at the time of their conception, semiconductor processes did not allow
close control of many circuit parameters, and designs were dimensioned to
accommodate that. Newer designs rely on much narrower control of circuit
parameters. They still drift under irradiation however, so these designs
may have very poor radiation hardness.

\subsubsection{Dynamic effects of radiation}
Radiation can trigger brief current impulses in semiconductors, which
may change the state of flip-flops or memory cells, so-called Single
Event Upsets. The lesson is to avoid relying on state held in logic
subject to radiation. Some remedial techniques might be to rewrite the
logic state regularly from a remote location, or to use redundancy and
error correction logic. For analog circuitry, there is really no
other solution than to move it away.

Many integrated circuits can suffer from latch-up. This often involves
parasitic components that aren't shown in the device schematics, if such
schematics are available at all. Latch-up is often destructive unless
the circuit is somehow protected. This protection could take the form
of compartmentalized and over-current-protected power distribution,
possibly with latch-up detection and remediation circuitry.

Discrete designs are more robust against latch-up, because the parasitic
elements that fragilize integrated circuits are absent in discrete
components. ICs can be hardened by using semiconductor-on-insulator
(SOI) techniques, which eliminate most of the troublesome parasitic
circuit elements. Of course, these techniques are more costly.

\section{Conclusion}

Analog electronics has an important role to play in beam
instrumentation. An instrument with a well-designed analog front-end 
will out-perform anything where this issue has been neglected.
The problem facing beam instrumentation designers is the optimal 
extraction of a useful signal, which involves filtering, impedance
and noise matching, amplification, 
transmission of signals over distances from decimeters
to hectometers, rejection of interference, signal integrity,
and radiation effects. Hopefully I have contributed something
useful on these subjects.

\end{document}